\documentclass[aps,superscriptaddress,nofootinbib,showpacs,eqsecnum,preprint,tightenlines]{revtex4}
\usepackage{hyperref}
\usepackage{epsfig,rotating}
\usepackage{amsmath,amssymb}
\usepackage{dsfont}
\usepackage{bbm}
\usepackage{slashed}
\numberwithin{equation}{section}

\newcommand{\be}{\begin{equation}}
\newcommand{\ee}{\end{equation}}
\newcommand{\bea}{\begin{eqnarray}}
\newcommand{\eea}{\end{eqnarray}}

\newcommand{\vp}{\vec{p}}

\newcommand{\vq}{\vec{q}}

\newcommand{\vk}{\vec{k}}

\newcommand{\Mpi}{M_{\pi \rightarrow \phi_1 \phi_2}}
\newcommand{\Mfi}{M_{\phi_1 \rightarrow \chi_1 \chi_2}}

\newcommand{\epsfi}{E_{\phi\phi}}
\newcommand{\epsfichi}{E_{\phi\chi\chi}}

\begin{document}

\title{Time evolution of cascade decay.}

\author{Daniel Boyanovsky}
\email{boyan@pitt.edu}
\author{Louis Lello}
\email{lal81@pitt.edu}
\affiliation{Department of Physics and
Astronomy, University of Pittsburgh, Pittsburgh, PA 15260}

\date{\today}

\begin{abstract}
 We study non-perturbatively the   time evolution of cascade decay for generic fields $\pi \rightarrow \phi_1\phi_2\rightarrow \phi_2\chi_1\chi_2$ and obtain the time dependence of  amplitudes and populations for the resonant and final states. We  analyze in detail the different time scales  and the manifestation of unitary time evolution   in the dynamics of production and decay of resonant intermediate and final states. The probability of occupation (population) ``flows'' as a function of time from the initial to the final states. When the decay width of the parent particle $\Gamma_\pi$  is much larger than that of the intermediate resonant state  $\Gamma_{\phi_1}$  there is a ``bottleneck'' in the flow, the population of resonant states builds up to a maximum at $t^* = \ln[\Gamma_\pi/\Gamma_{\phi_1}]/(\Gamma_\pi-\Gamma_{\phi_1})$
  nearly saturating unitarity and decays to the final state on the longer time scale $1/\Gamma_{\phi_1}$. As a consequence of the wide separation of time scales  in this case  the cascade decay can be interpreted as evolving sequentially $\pi \rightarrow \phi_1\phi_2; ~ \phi_1\phi_2\rightarrow \phi_2\chi_1\chi_2$. In the opposite limit the population of resonances ($\phi_1$) does not build up substantially and the cascade decay proceeds almost directly from the initial parent  to the final state without resulting in a large amplitude of the resonant state. An alternative but equivalent non-perturbative method useful in cosmology is presented.  Possible phenomenological implications for heavy sterile neutrinos as resonant states and consequences of quantum entanglement and correlations in the final state are discussed.

\end{abstract}


\maketitle

\section{Introduction}
The   decay of unstable or metastable states via a cascade $A \rightarrow BC\rightarrow BXY$ is of interdisciplinary  importance  in particle physics, quantum optics and cosmology. Early studies in particle physics proposed to use the time evolution of intermediate (resonant) states in cascade decays of heavy  (B) mesons to study aspects of CP violation, mixing phenomena\cite{sanda,azimov,azidun,kayser,silva,grimus} and CPT violation\cite{lavoura,nobary}.
The Belle collaboration\cite{belle} has reported on remarkably precise measurements of (EPR) entanglement and correlations in cascade decays  $\Upsilon(4S)\rightarrow B^{0}\overline{B}^{0}\rightarrow (l,J/\Psi,K)$  via the analysis of the time dependence of the flavor asymmetry\cite{yab}. The BaBar collaboration\cite{trevbo,babar} has reported on the first direct measurement of time reversal violation in the  $B^{0}\overline{B}^{0}$ system from $\Upsilon(4S)$ decay at rest by studying the time dependent correlations between the members of the entangled $B^{0}\overline{B}^{0}$ pairs also in a cascade decay.

More recently cascade decays   have been proposed as possible mechanisms of CP and lepton flavor violation mediated by heavy sterile neutrinos as resonant states\cite{cvetic} and as possible solutions to the LSND/MiniBooNe anomalies\cite{gninenko}. Cascade decays    may probe new particles beyond the standard model via kinematic edges associated with the new degrees of freedom in the intermediate states\cite{grossman}.

In quantum optics and    cavity Quantum Electrodynamics, the cascade decay of multi-level atoms or quantum dot systems  is studied as a source of correlated photons and entanglement\cite{qobooks1,qobooks2,qobooks3,garraway,troiani,qd} with potential applications in quantum information. The possibility of observation of polarization entangled photon pairs in cascade decay in quantum dots has recently been advanced, such a measurement is, fundamentally, similar to the observation of CP violating amplitudes in the time evolution of a cascade decay in B-mesons and similar meson systems addressed in refs.\cite{sanda}-\cite{grimus}.

In inflationary cosmology the rapid expansion of the Universe entails that there is no time-like Killing vector and as a consequence of the lack of  kinematic thresholds   quanta of a field can decay into quanta of the \emph{same} field \cite{woodard,boydecay,wwds,marolf,akhmedov} leading to the recent suggestion of  cascade decay of inflaton quanta during inflation\cite{lellods} and the concomitant kinematic entanglement of the produced particles.

\vspace{1mm}

\textbf{Motivation and Goals:}
In the analysis of cascade decays in refs.\cite{sanda,azimov,azidun,kayser,silva}, the time evolution of the cascade is analyzed as a \emph{sequential} series of events. Consider an initial state $I$ decaying into an intermediate state $M$, which in turn decays into a final state $f$, the amplitude for such process is proposed to be
\be A[I \stackrel{t_I}{\rightarrow}M \stackrel{t_M}{\rightarrow}f] = e^{-iW_I\,t_I}~e^{-iW_M\,t_M}~\mathcal{M}_{I\rightarrow M}~\mathcal{M}_{M\rightarrow f} \label{transitions}\ee where $t_I$ is the time at which the initial state $I$ decays into the intermediate resonant state $M$ and $t_M$ is the time at which the intermediate state $M$ decays into the final state in their respective rest frames,
$W_{I,M} = m_{I,M} -i\Gamma_{I,M}$ are the complex energies of the corresponding states and $\mathcal{M}$ are the corresponding transition amplitudes (we neglect here the possibility of mixing in the initial, intermediate or final states).

While this sequential characterization  may be phenomenologically suitable to the description of experimental situations in which there is a wide separation in time scales as a consequence of large differences in widths and masses, it clearly \emph{assumes} that the decays occur at specific (proper) times $t_I,t_M$ in   sequence.

However, in a general case, this   is at odds with the description of radiative cascades in multi-level systems in quantum optics\cite{qobooks1,qobooks2,qobooks3} and with the description of decay of an unstable (or metastable) state as a continuous process with a (generally exponential) time distribution in which the amplitude for the parent particle decays exponentially and the amplitude of the daughter (resonant state) increases continuously on a similar time scale.

The interdisciplinary relevance of cascade decay motivates us to study the time evolution within a framework that is suitable to extension to the realm of cosmology and that could prove useful in other areas such as quantum optics and condensed matter physics.

The goal of this article is to study the full  time evolution of a cascade decay process directly from the quantum mechanical evolution of an initial quantum state. In particular we address the important issue of unitarity in the time evolution from the initial to the final state, focusing  on the time evolution of the amplitudes  and populations of the intermediate resonant and final states and how unitarity is manifest in these amplitudes.

 For this purpose we provide a non-perturbative quantum field theoretical generalization of the methods ubiquitous in quantum optics\cite{qobooks1,qobooks2,qobooks3,garraway} adapted to the realm of a full quantum field theory.  Furthermore, one of our objectives is to provide a non-perturbative method to study the time evolution directly in real time that would be suitable for applications in cosmology where the expansion of the Universe introduces an explicit time dependence in the evolution Hamiltonian\cite{boydecay,akhmedov,lellods}.

 In this article we consider  the case of only one resonant intermediate state to introduce and develop the methods and to exhibit the main physical processes  in a simpler setting, postponing the study of mixing and oscillations in the resonant state for future study.

 \vspace{2mm}

\textbf{Brief summary of results:}
We implement a non-perturbative method that yields the full time evolution of an initial state of a decaying parent particle in a model of generic fields that incorporates the main features of a cascade decay via a resonant state $\pi \rightarrow \phi_1\phi_2\rightarrow \phi_2\chi_1\chi_2$  although the results are general. We obtain the time evolution of the amplitudes for the intermediate and final states and show explicitly that the method is akin to a Dyson-type resummation of self-energies and fulfills unitarity. Unitary time evolution is manifest in a transfer or ``flow'' of probability or population from the initial through the intermediate resonant to the final state. We analyze in detail the evolution of the populations on the different  time scales. The population of the resonant state grows at early times reaching a maximum at $t^* = \ln[\Gamma_\pi/\Gamma_{\phi_1}]/(\Gamma_\pi-\Gamma_{\phi_1})$ and decaying to the final state.

The build-up of the population of the intermediate resonant state depends crucially on the ratio of decay widths $\Gamma_\pi/\Gamma_{\phi_1}$. For $\Gamma_\pi/\Gamma_{\phi_1}\gg 1$ there is a ``bottleneck'' in the sense that the population of the resonant state builds up to nearly saturate unitarity   on a time scale $t\simeq t^*$ and decays into the final state on much longer time scales $\simeq 1/\Gamma_{\phi_1}$. In the opposite limit, the population of the initial state is transferred almost directly to the final state with a very small build-up of the population of the intermediate resonant state.

 The final asymptotic state after both the parent particle and intermediate resonant state decays is a many particle   state featuring quantum entanglement and correlations among the final particles.

We also provide a quantum field theoretical  generalization of the   Wigner-Weisskopf method that provides a similar type of non-perturbative resummation directly in real time and yields the same results but is amenable of implementation in cosmology where the interaction Hamiltonian is explicitly dependent on time.

We conjecture on possible phenomenological consequences in particular for heavy sterile neutrinos as intermediate resonant states in pseudoscalar  decays.

\section{The model}\label{sec:model}
We consider a model of generic real scalar fields $\pi,\phi_{1,2},\chi_{1,2}$ of  masses $M_\pi;m^\phi_{1,2},m^\chi_{1,2}$ respectively to study the relevant phenomena in the simplest setting and to focus on the main aspects of the method and physical processes, however the main results will be argued  to be general.

The total Hamiltonian is $H=H_0+H_I$ with $H_0$ the free field Hamiltonian and
\be H_I = \int d^3 x \Big\{g_\pi\,\pi(x)\phi_1(x)\phi_2(x)+ g_\phi\, \phi_1(x)\chi_1(x)\chi_2(x) \Big\}\,.  \label{HI} \ee

Let us consider an initial state   with one $\pi$ particle of momentum $\vec{k}$ and the vacuum for the other fields, namely
\be \big|\Psi(t=0)\rangle = \big|\pi_{\vec{k}} \rangle \,. \label{inistate} \ee

\begin{figure}[h!]
\begin{center}
\includegraphics[height=3in,width=3in,keepaspectratio=true]{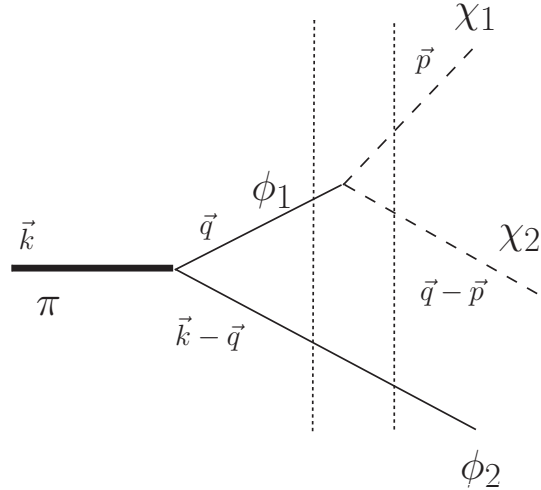}
\caption{Cascade decay $\pi \rightarrow \phi_1\phi_2; \phi_1\rightarrow \chi_1\chi_2$.The dashed lines depict the intermediate two particle state and the final three particle state.  }
\label{fig:cascade}
\end{center}
\end{figure}

Upon time evolution this state evolves into $\big|\Psi(t)\rangle$ obeying
\be \frac{d}{dt} \big|\Psi(t)\rangle = -i (H_0+H_I) \big|\Psi(t)\rangle\,, \label{timeevol} \ee
when $m_\pi > m_{\phi_1}+m_{\phi_2}~;~m_{\phi_1} > m_{\chi_1}+m_{\chi_2}$ the interaction Hamiltonian (\ref{HI}) describes the cascade process depicted in fig.\ref{fig:cascade}.

At any given time the state $\big|\Psi(t)\rangle $ can be expanded in Fock states of the non-interacting Hamiltonian $H_0$
\be \big|\Psi(t)\rangle = \sum_{n}C_n(t) \,\big|n\rangle ~~;~~H_0 \big|n\rangle = E_n \big|n\rangle \label{freexp}\ee the time evolution of the coefficients $C_n(t)$ is obtained from (\ref{timeevol}), and projecting onto the states $\big|n\rangle$ namely
\be \dot{C}_n(t)= -iE_n\,C_n(t)+ \sum_{\kappa} \langle n\big|H_I\big|\kappa \rangle\, C_\kappa(t) \,. \label{eqns}\ee This is an infinite hierarchy of equations, progress is made by truncating the hierarchy at a given order and solving the coupled equations, this method is equivalent to a \emph{Dyson resummation of self-energy diagrams} as will be seen clearly  and in detail   below.

 The time evolution of the initial state depicted up to second order in the interaction in fig.\ref{fig:cascade} leads to considering the following form
\be \big|\Psi(t)\rangle = C_\pi(\vec{k},t)\big|\pi_{\vec{k}}\rangle + \sum_{\vq}\,C_{\phi\phi}(\vk,\vq;t)\,\big|\phi_{1,\vq}\phi_{2,\vk-\vq} \rangle + \sum_{\vq;\vp}\,C_{\phi\chi\chi}(\vk,\vq,\vp;t)\,\big|\phi_{2,\vk-\vq}\chi_{1,\vp}\chi_{2,\vq-\vp} \rangle  +\cdots \label{state} \ee where the dots stand for many particle states that emerge in higher order in $H_I$. The matrix elements of the interaction Hamiltonian in (\ref{eqns}) describe transitions between single and multiparticle states, and the functions $C_\pi(\vec{k},t);C_{\phi\phi}(\vk,\vq;t); C_{\phi\chi\chi}(\vk,\vq,\vp;t)$ represent the amplitudes of the initial single particle state and  multiparticle states in the time evolved state. The initial conditions on these amplitudes are
\be C_\pi(\vec{k},0)=1~;~C_{\phi\phi}(\vk,\vq;0)=0~;~C_{\phi\chi\chi}(\vk,\vq,\vp;0)=0 \,. \label{iniconds}\ee

The evolution equations for the amplitudes are obtained by projection, as in (\ref{eqns}). Introducing the shorthand definitions
\bea \Mpi(\vk,\vq) & \equiv &  \langle \pi_k|H_I|\phi_{1,\vq}\,\phi_{2,\vk-\vq}\rangle \label{Mpi}\\
\Mfi(\vq,\vp) & \equiv &  \langle \phi_{1,\vq}|H_I|\chi_{1,\vp}\,\chi_{2,\vq-\vp}\rangle \label{Mfi}\eea and

\bea
E_{\phi \phi}(\vk,\vq) & \equiv & E^{\phi_1}_{\vq}+E^{\phi_2}_{\vk-\vq} \label{epsfi} \\
E_{\phi \chi\chi}(\vk,\vq,\vp) & \equiv & E^{\phi_2}_{\vk-\vq}+E^{\chi_1}_{\vp}+E^{\chi_2}_{\vq-\vp} \label{epsfichi} \eea  where $E_{\vk}$ are the single particle energies for $\pi,\phi,\chi$ respectively we find
\be    \dot{C}_\pi(\vec{k},t)   =    -iE^\pi_k\, C_\pi(\vec{k},t)-i \sum_{\vq}\,\Mpi(\vk,\vq) \,C_{\phi\phi}(\vk,\vq;t)\label{dotcpi}\ee
\bea \dot{C}_{\phi\phi}(\vk,\vq;t)  &  = & -i E_{\phi \phi}(\vk,\vq) \, {C}_{\phi\phi}(\vk,\vq;t)-i \Mpi^*(\vk,\vq) \, {C}_\pi(\vec{k},t) \nonumber \\ && -i \sum_{\vp} \Mfi (\vq,\vp)\, C_{\phi\chi\chi}(\vk,\vq,\vp;t) \label{dotcfifi}\eea
\bea \dot{C}_{\phi\chi\chi}(\vk,\vq,\vp;t) & = & -iE_{\phi \chi \chi}(\vk,\vq,\vp)\, {C}_{\phi\chi\chi}(\vk,\vq,\vp;t)\, -i \Mfi^*(\vq,\vp)\,{C}_{\phi\phi}(\vk,\vq;t) \label{dotcfichi} \eea We have truncated the hierarchy of equations up to second order in the interaction Hamiltonian (\ref{HI}), thereby neglecting the higher order branches with four particles etc. As it will be demonstrated below, the solution of the coupled hierarchy of equations up to this order provides   Dyson resummations of \emph{both} the propagator of the $\pi$ particle \emph{and} the propagator for the $\phi_1$ particle which corresponds to the resonant intermediate state in the cascade decay process. This will become clear below.

In what follows we will suppress the arguments of the various functions introduced above to simplify notation.

It proves convenient to solve these coupled set of equations by Laplace transform as befits an initial value problem. Defining the Laplace transform of $C(t)$ as
\be \tilde{C}(s) = \int_0^{\infty}e^{-st} C(t) dt \label{lapla} \ee and with the initial conditions (\ref{iniconds}) the set of coupled equations  (\ref{dotcpi}-\ref{dotcfichi}) becomes
\be \tilde{C}_\pi(\vk,s)\,[s+iE^\pi_k] = 1-i\sum_{\vq}\Mpi(\vk,\vq)\, \tilde{C}_{\phi\phi}(\vk,\vq;s)\,, \label{laplacpi}\ee
\bea
 \tilde{C}_{\phi\phi}(\vk,\vq;s)[s+i\epsfi] & = & -i \Mpi^*(\vk,\vq) \, \tilde{C}_{\pi}(\vk;s) \nonumber \\ && -i\sum_{\vp}\Mfi(\vq,\vp)\,\tilde{C}_{\phi\chi\chi}(\vk,\vq,\vp;s) \,, \label{laplacfi}\eea
\be \tilde{C}_{\phi\chi\chi}(\vk,\vq,\vp;s)[s+i\epsfichi] = -i \Mfi^*(\vq,\vp)\, \tilde{C}_{\phi\phi}(\vk,\vq;s) \,.\label{laplacfichi}\ee We now solve this system of algebraic equations from the bottom up obtaining,
\be \tilde{C}_\pi(\vk;s) = \frac{1}{s+iE^\pi_k+i\Sigma^\pi(\vk;s)}\,, \label{tilCpi}\ee
\be \tilde{C}_{\phi\phi}(\vk,\vq;s) = -i\frac{\Mpi^*\,\tilde{C}_\pi(\vk;s)}{s+i\epsfi +i\Sigma^\phi(\vk;s)} \,, \label{tilCfi}\ee
\be \tilde{C}_{\phi\chi\chi}(\vk,\vq,\vp;s) = -i\frac{\Mfi^*\,\tilde{C}_{\phi\phi}(\vk;s)}{s+i\epsfichi}\,, \label{tilCfichi1}\ee where the self-energies are given by
\be i\Sigma^\pi(\vk;s) = \sum_{\vq} \frac{|\Mpi|^2}{s+i\epsfi+i\Sigma^\phi(\vk,\vq;s)}\,,\label{sigmapi}\ee and

\be i\Sigma^\phi(\vk,\vq;s) = \sum_{\vp} \frac{|\Mfi|^2}{s+i\epsfichi}\,.\label{sigmafi}\ee

\begin{figure}[h!]
\begin{center}
\includegraphics[height=3.5in,width=4in,keepaspectratio=true]{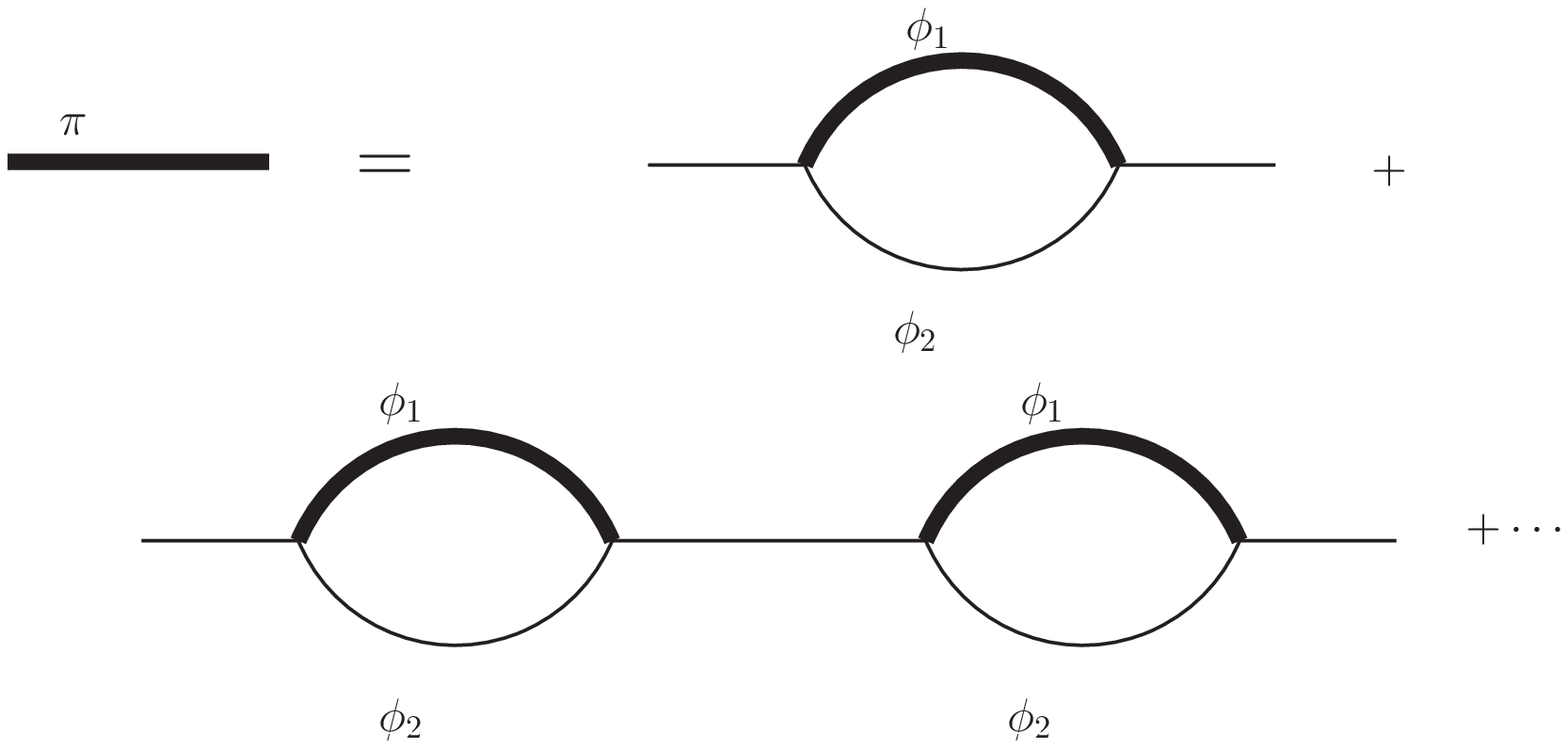}\\
\includegraphics[height=3.5in,width=4in,keepaspectratio=true]{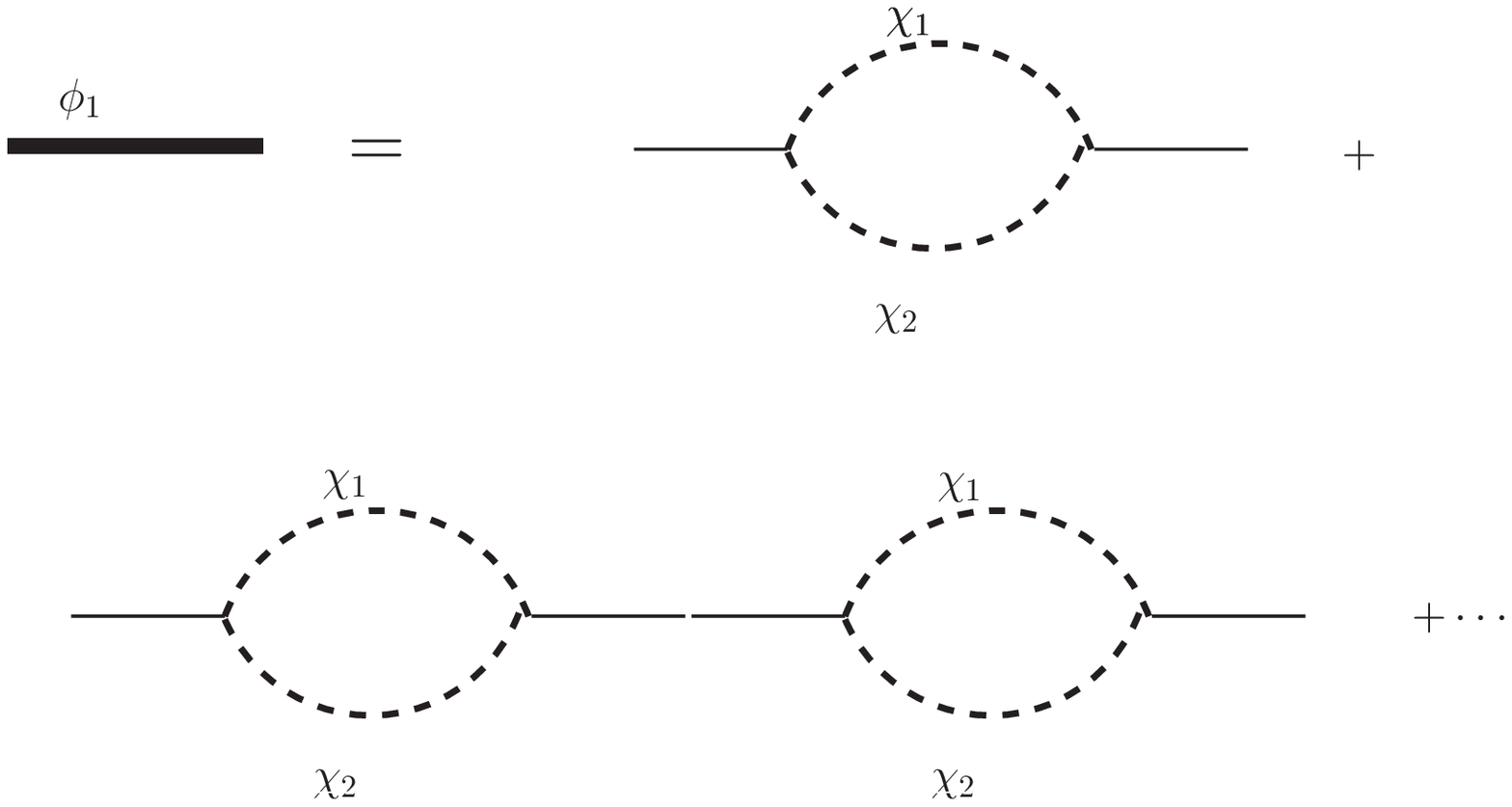}
\caption {$\pi$ propagator, the thick $\phi_1$ line is the full $\phi_1$ propagator with self-energy resummation.  }
\label{fig:piprop}
\end{center}
\end{figure}

These equations have a familiar interpretation in terms of Dyson resummations of self-energy diagrams depicted in fig.(\ref{fig:piprop}): the irreducible self-energy for the $\pi$-field is in terms of the full propagator of the (resonant) $\phi_1$ particle which is itself given by a Dyson resummation of irreducible self-energy diagrams involving a loop of the final state particles. It is at this stage that the equivalence between the truncation of the hierarchy of equations and the resummation in terms of a Dyson series becomes manifest. Therefore, the solution of the truncated hierarchy up to the given (second) order, provides a non-perturbative resummation of self-energy diagrams both for the decaying parent particle \emph{and} the intermediate resonant state. The $\pi$ self-energy (\ref{sigmapi}) includes the self-energy correction to the (intermediate) $\phi_1$ state.

\vspace{2mm}

\subsection{Time evolution:}
The time evolution is obtained by performing the inverse Laplace transform:
\be C(t) = \int_{\mathcal{C}} \frac{ds}{2\pi\,i}~ e^{st} ~\tilde{C}(s) \label{antilapla}\ee where  $\mathcal{C}$ stands for the Bromwich contour parallel to the imaginary axis in the complex $s$ plane and to the right of all the singularities of the function $\tilde{C}(s)$.

With the purpose of using the convolution theorem
\be \int_{\mathcal{C}} \frac{ds}{2\pi\,i}~ e^{st} ~\tilde{F}(s)\,\tilde{G}(s) = \int_0^t F[t-t']\,G[t'] dt' \label{convo}\ee where $F[t],G[t]$ are the inverse Laplace transforms of $\tilde{F}(s),\tilde{G}(s)$ respectively,
it proves convenient to rewrite (\ref{tilCfi},\ref{tilCfichi1}) as

 \be \tilde{C}_{\phi\phi}(\vk,\vq;s) = -i\Mpi^*\,\tilde{C}_\pi(\vk;s) ~\tilde{G}_{\phi\phi}(s) ~~;~~\tilde{G}_{\phi\phi}(s)=
 \frac{1}{s+i\epsfi +i\Sigma^\phi(s)} \label{Gfi}\ee

\be \tilde{C}_{\phi\chi\chi}(\vk,\vq,\vp;s) = -i \Mfi^*\,\tilde{C}_{\phi\phi}(\vk;s)\,\tilde{G}_{\phi\chi\chi}(s) ~~;~~\tilde{G}_{\phi\chi\chi}(s)=
\frac{1}{s+i\epsfichi} \label{tilCfichi2}\ee

The amplitudes $\tilde{C}(s)$ given by eqns. (\ref{tilCpi}-\ref{tilCfichi1}) generally feature branch cuts above multiparticle thresholds and complex poles   with $\mathrm{Re}s \leq 0$ namely either along the imaginary axis or to its left as befits stable particles, multiparticle cuts (along the imaginary axis) or decaying resonances. Therefore the Bromwich contour corresponds to integration along the path parallel to the imaginary axis $s = i\omega + \varepsilon~;~ -\infty \leq \omega \leq \infty~;~\varepsilon \rightarrow 0^+$ and
\be C(t) = \int^\infty_{-\infty}\frac{d\omega}{2\pi} \,e^{i\omega t}\,\tilde{C}(s=i\omega+\varepsilon) \label{rtime}\ee where the contour must now be closed in the upper half $\omega-$ plane.

The first step in obtaining the time evolution of the amplitudes is to obtain $C_\pi(t)$ for which we need to identify the singularities in

\be\tilde{C}_\pi(s=i\omega+\varepsilon) = \frac{1}{i\omega+iE^\pi_k+i\Sigma^{\pi}(\vk,i\omega+\varepsilon)+\varepsilon}\,. \label{Cpiome}\ee
 The self-energy features a two-particle cut and the $\pi$ particle becomes a resonance if $E^\pi_k$ is embedded in the two-particle continuum. For a weakly coupled theory this resonance is described by a complex pole in the upper half plane very near the real axis, since in absence of perturbations the pole is at $\omega = -E^\pi_k$. Consistently in perturbation theory we write $\omega = -E^\pi_k$ in the argument of $\Sigma^\pi$, with
 \be i\Sigma^{\pi}(\vk,-iE^\pi_k+\varepsilon) =  \sum_{\vq} \frac{|\Mpi|^2}{-iE^\pi_k+i\epsfi+i\Sigma^\phi(\vk,\vq;-iE^\pi_k+\varepsilon)+\varepsilon} \label{sigpiome}\ee and from (\ref{sigmafi}) we find
 \be i\Sigma^\phi(\vk,\vq;s=-iE^\pi_k+\varepsilon) = i\sum_{\vp} \frac{|\Mfi|^2}{(E^\pi_k-\epsfichi)+i\varepsilon} = i\delta E + \frac{1}{2}\,\gamma(\vk,\vq) \label{sigmafiome}\ee
 \bea \delta E & = &   \sum_{\vp}\mathcal{P}~ \frac{|\Mfi|^2}{(E^\pi_k-\epsfichi)}\,,\label{shiftEp}\\
\gamma(\vk,\vq) & = & 2\pi \sum_{\vp}  |\Mfi|^2~\delta(E^\pi_k-\epsfichi)\,, \label{gama}\eea we note that $\delta E$ is \emph{not} the renormalization of the $\phi_1$ energy (mass) and $\gamma(\vk,\vq)$ is \emph{not} its decay width, because in the denominator of (\ref{shiftEp}) and the argument of the delta function in (\ref{gama}) is $E^\pi_k- E^{\phi_2}_{\vk-\vq}-E^{\chi_1}_{\vp}-E^{\chi_2}_{\vq-\vp} $ instead of being $E^{\phi_1}_{\vq}-E^{\chi_1}_{\vp}-E^{\chi_2}_{\vq-\vp}$ (see below for clarification on this point).

Inserting  this expression into (\ref{sigpiome}) we find
\be i\Sigma^{\pi}(\vk,-iE^\pi_k+\varepsilon) = i\Delta E^\pi +\frac{\Gamma_\pi(k)}{2} \label{sigpi} \ee
  where the energy shift
\be \Delta E^\pi = \sum_{\vq} \frac{|\Mpi|^2~(E^\pi_k- \epsfi-\delta E)}{\Big(E^\pi_k- \epsfi-\delta E\Big)^2+\Big(\frac{\gamma(\vk,\vq)}{2}\Big)^2} \label{DeltaE}\ee   is absorbed into  a renormalization of the $\pi$ mass, and
\bea \Gamma_\pi(k)  & = &   \sum_{\vq} \frac{|\Mpi|^2~\gamma(\vk,\vq)}{\Big(E^\pi_k- \epsfi-\delta E\Big)^2+\Big(\frac{\gamma(\vk,\vq)}{2}\Big)^2} \nonumber \\
& = & 2\pi\sum_{\vq}  \sum_{\vp} \frac{|\Mpi|^2~|\Mfi|^2~\delta(E^\pi_k-\epsfichi)}{\Big(E^\pi_k- \epsfi-\delta E\Big)^2+\Big(\frac{\gamma(\vk,\vq)}{2}\Big)^2}
 \label{gamapi}\eea is the decay width of the $\pi$ particle.

As mentioned above  $ \gamma(\vk,\vq)$ is \emph{not} the width of the $\phi_1$ resonance, and (\ref{shiftEp}) is \emph{not} the renormalization of the $\phi_1$ energy (mass renormalization), however in perturbation theory    $\Gamma_\pi$ (\ref{gamapi}) becomes resonant as $E^\pi_k \rightarrow E_{\phi\phi}$, and the sum is dominated by this resonance. Near this resonance one can replace $E^\pi \rightarrow E_{\phi \phi}$ into (\ref{shiftEp},\ref{gama}) and recognizing from (\ref{epsfi},\ref{epsfichi}) that $E_{\phi\phi}-E_{\phi\chi\chi} = E^{\phi_1}_{\vq}-E^{\chi_1}_{\vp}-E^{\chi_2}_{\vq-\vp} $ it follows that \emph{ near this resonance }
\be \delta E \rightarrow \Delta E^{\phi_1} = \sum_{\vp}\mathcal{P}~ \frac{|\Mfi|^2}{(E^{\phi_1}_{\vq}-E^{\chi_1}_{\vp}-E^{\chi_2}_{\vq-\vp} )}\label{deltaEfi1}\ee
 is the energy shift absorbed into a  renormalization of the $\phi_1$ mass and
  \be \gamma(\vk,\vq)\rightarrow \Gamma_{\phi_1}(\vq) = 2\pi \sum_{\vp}  |\Mfi|^2~\delta(E^{\phi_1}_{\vq}-E^{\chi_1}_{\vp}-E^{\chi_2}_{\vq-\vp} ) \,,\label{Gammafi1}\ee is  the decay width of $\phi_1$.

  Therefore in a cascade decay where the intermediate $\phi_1$ becomes resonant (near on-shell) and absorbing $\delta E$ near this resonance into the renormalization of the  $\phi_1$ energy (mass), the $\pi$ decay rate (\ref{gamapi}) becomes
  \be  \Gamma_\pi(k)    \simeq  2\pi\sum_{\vq}  \sum_{\vp} \frac{|\Mpi|^2~|\Mfi|^2~\delta(E^\pi_k-E^{\phi_2}_{\vk-\vq}-E^{\chi_1}_{\vp}-E^{\chi_2}_{\vq-\vp} )}{\Big(E^\pi_k- E^{\phi_2}_{\vk-\vq}- E^{\phi_1}_{\vq} \Big)^2+\Big(\frac{\Gamma_{\phi_1}(\vq)}{2}\Big)^2}
 \label{gamapires}\ee  This is the usual expression of a decay rate of the parent particle in a cascade   in the Breit-Wigner approximation for the propagator of the resonant intermediate state.

 Furthermore in the narrow width approximation, (\ref{gamapires}) is dominated by the resonance in the denominator and to leading order in the width we can replace
 \be \frac{1}{\Big(E^\pi_k- E^{\phi_2}_{\vk-\vq}- E^{\phi_1}_{\vq} \Big)^2+\Big(\frac{\Gamma_{\phi_1}(\vq)}{2}\Big)^2} \rightarrow \frac{2\pi}{\Gamma_{\phi_1}(\vq)}\,\delta\Big(E^\pi_k- E^{\phi_2}_{\vk-\vq}- E^{\phi_1}_{\vq} \Big) \label{narrow}\ee and write (\ref{gamapires}) as
 \be \Gamma_\pi(k)    \simeq  2\pi\sum_{\vq}    |\Mpi|^2 \, \delta\Big(E^\pi_k- E^{\phi_2}_{\vk-\vq}- E^{\phi_1}_{\vq} \Big)\, \Bigg[\frac{2\pi}{\Gamma_{\phi_1}(\vq)} \sum_{\vp} |\Mfi|^2 \, \delta(E^{\phi_1}_{\vq} -E^{\chi_1}_{\vp}-E^{\chi_2}_{\vq-\vp} )\Bigg]\,. \label{gamapifini}\ee From (\ref{Gammafi1}) we see that the bracket in this expression equals one, leading to
 \be \Gamma_\pi(k)    \simeq  2\pi\sum_{\vq}    |\Mpi|^2 \, \delta\Big(E^\pi_k- E^{\phi_2}_{\vk-\vq}- E^{\phi_1}_{\vq} \Big) \label{gamapnw}\ee which is valid in the narrow width approximation. In this case there is no other decay channel for the resonant state and the bracket in (\ref{gamapifini}) becomes unity, however if there are other decay channels this bracket would be replaced by the branching ratio $BR(\phi_1\rightarrow \chi_1\chi_2)$, which is the usual result for resonant decay in the narrow width approximation.

We are now in position to obtain the time evolution of the amplitudes. In the narrow width limit, the amplitude $\tilde{C}_\pi(\omega)$ given by (\ref{Cpiome})  features a (narrow) resonance near $\omega \simeq -E^\pi_k$  and is of the Breit-Wigner form
  \be\tilde{C}_\pi( \omega\simeq -E^\pi_k) \simeq \frac{-i}{ \omega + E^\pi_k  -\frac{i}{2}\Gamma_{\pi}(\vk)} \label{Cpbw}\ee where now $E^\pi_k$ is the renormalized $\pi$ single particle energy and $\Gamma_\pi$ is the decay width. This is equivalent to a Breit-Wigner approximation to the propagator in terms of a complex pole and as usual ignores the ``background'' contribution which is perturbative. The time evolution of the amplitude follows straightforwardly, it is given by
  \be C_\pi(\vk,t) = e^{-iE^\pi_k\,t}\,e^{-\frac{\Gamma_\pi(k)}{2}\,t} \,.\label{cpioft}\ee

 In a similar manner we now obtain the time evolution of $G_{\phi\phi}(t)$, the anti Laplace transform of $\tilde{G}_{\phi\phi}(s)$ in eqn. (\ref{Gfi}),  with
 \be \tilde{G}_{\phi\phi}(s=i\omega+\varepsilon) =
 \frac{1}{i\omega+i\epsfi +i\Sigma^\phi(\vk;i\omega+\varepsilon)+\varepsilon} \,.\label{Gfiomega}
 \ee It features a  complex  pole near $\omega \simeq -\epsfi$ with
 \be i \Sigma^\phi(s= -i\epsfi+\varepsilon) = i\Delta E^{\phi_1} + \frac{\Gamma_{\phi_1}}{2} \label{sigfi1}\ee where the energy shift $ \Delta E^{\phi_1}$  and decay width $\Gamma_{\phi_1}$ are given by
 (\ref{deltaEfi1},\ref{Gammafi1}) respectively, and the energy shift $ \Delta E^{\phi_1}$ is   absorbed into a renormalization of the $\phi_1$ mass.

 As in the case of $C_\pi(t)$ we now obtain
 \be G_{\phi\phi}(t) = e^{-i\epsfi\,t}\,e^{-\frac{\Gamma_{\phi_1}}{2}\,t}\,, \label{Gfifit}\ee where $E_{\phi_1}$ in $\epsfi$ is the renormalized single particle energy for $\phi_1$.  We now implement the convolution theorem (\ref{convo}) with $C_\pi(t),G_{\phi\phi}(t)$ given by (\ref{cpioft},\ref{Gfifit}) respectively and find
 \be C_{\phi\phi}(t) = \Mpi^*\, \frac{\Big[ e^{-iE^\pi_k\,t}\,e^{-\frac{\Gamma_\pi(k)}{2}\,t}- e^{-i\epsfi\,t}\,e^{-\frac{\Gamma_{\phi_1}}{2}\,t}\Big]}{\Big(E^\pi-\epsfi\Big)-\frac{i}{2}\Big(\Gamma_\pi-\Gamma_{\phi_1}\Big)} \,. \label{Cfifit}\ee This is the amplitude of the two particle intermediate state with a resonant $\phi_1$.

From eqn. (\ref{tilCfichi1},\ref{tilCfichi2}) and with the anti Laplace transform of $\tilde{G}_{\phi\chi\chi}$ given by
\be {G}_{\phi\chi\chi}(t) = e^{-i\epsfichi\,t} \label{antilapgifichi}\ee
 we find
\be C_{\phi\chi\chi}(t) = -i\Mfi^* \int^t_0 C_{\phi\phi}(t') \, e^{-i\epsfichi(t-t')} ~dt'\label{Cfichi2}\ee which yields (suppressing all the momenta labels)
\bea   C_{\phi\chi\chi}(t)   = && \frac{\Mpi^*\, \Mfi^* \,e^{-i\epsfichi\,t}}{\Big(E^\pi-\epsfi\Big)-\frac{i}{2}\Big(\Gamma_\pi-\Gamma_{\phi_1}\Big)}\times\nonumber \\
&& \Bigg\{\frac{\Big[e^{-i(E^\pi-\epsfichi)t}\,e^{-\frac{\Gamma_\pi}{2}t}-1 \Big] }{(E^\pi-\epsfichi)-i\frac{\Gamma_\pi}{2}}- \frac{\Big[e^{-i(\epsfi-\epsfichi)t}\,e^{-\frac{\Gamma_{\phi_1}}{2}t}-1 \Big] }{(\epsfi-\epsfichi)-i\frac{\Gamma_{\phi_1}}{2}} \Bigg\}\,.  \label{Cfichifin}\eea This is the  amplitude of the final three particle state. Although the amplitudes (\ref{Cfifit},\ref{Cfichifin}) look unfamiliar, it will be proven in the next section that they   satisfy unitarity.

\section{Unitarity: Population flow.}

Unitary time evolution of the state $|\Psi(t)\rangle$ (\ref{freexp}) with the initial condition (\ref{inistate}) implies $\langle \Psi(t)|\Psi(t) \rangle =1$, namely
\be \sum_{n}|C_n(t)|^2 = 1 \label{uni}\ee which for the state (\ref{state}) implies
\be   |C_\pi(\vec{k},t)|^2 + \sum_{\vq}\,|C_{\phi\phi}(\vk,\vq;t)|^2  + \sum_{\vq;\vp}|C_{\phi\chi\chi}(\vk,\vq,\vp;t)|^2  +\cdots = 1\,. \label{unitary} \ee

The number of $\pi$ particles with momentum $\vk$ is given by
\be   n^{\pi}_{\vk}(t) = \langle \Psi(t)|a^\dagger_{\pi,\vk}\,a_{\pi,\vk}|\Psi(t)\rangle = |C_{\pi}(\vk;t)|^2 \label{numberpi}\ee similarly the number   of   $\phi_1$ particles with particular momentum $\vq$ in the time evolved state $|\Psi(t)\rangle$ is given by
\be n^{\phi_1}_{\vq}(t) = \langle \Psi(t)|a^\dagger_{\phi_1,\vq}\,a_{\phi_1,\vq}|\Psi(t)\rangle = |C_{\phi\phi}(\vk,\vq;t)|^2 \label{numberfi1}\ee and the number of \emph{pairs} of $\chi_{1,2}$ particles with momenta $\vp,\vq-\vp$ respectively is
\be n^{\chi_1}_{\vp}(t)~n^{\chi_2}_{\vq-\vp}(t) = \langle \Psi(t)|\big(a^\dagger_{\chi_1,\vp}\,a_{\chi_1,\vp}\big)\big(a^\dagger_{\chi_2,\vq-\vp}\,a_{\chi_2,\vq-\vp}\big)|\Psi(t)\rangle =
|C_{\phi\chi\chi}(\vk,\vq,\vp;t)|^2 \label{pairchi}\ee where $a^\dagger_{\alpha}~;~a_{\alpha}$ are the annihilation and creation operators for the quanta of the respective fields. Therefore the probabilities $|C(t)|^2$ are also a measure of the \emph{population} of the many particle states upon decay of the initial state.

Since $|C_\pi(\vec{k},0)|  =1~;~ |C_{\phi\phi}(\vk,\vq;0)| =0 ~;~ |C_{\phi\chi\chi}(\vk,\vq,\vp;0)|=0$,   as time evolves the probabilities $|C_{\phi\phi}(\vk,\vq;t)|^2$ of the intermediate (resonant) state  $\big|\phi_{1,\vq}\phi_{2,\vk-\vq} \rangle $ will grow as the $|\pi\rangle$ state decays   therefore producing $\phi_1$ particles,   however these particles eventually decay  into  final state particles $\big|\phi_{2,\vk-\vq}\chi_{1,\vp}\chi_{2,\vq-\vp} \rangle $ whose population $|C_{\phi\chi\chi}(\vk,\vq,\vp;t)|^2 $ will grow in time . Therefore we expect the physical picture:  the amplitude of the initial state decays, while  the amplitude of the intermediate resonant state grows at early times   eventually decays into $\chi_{1,2}$ particles, and the amplitude of the final state grows more slowly than the intermediate state at early times, but it reaches an asymptotic final value  to fulfill the unitarity relation (\ref{unitary}) with $|C_\pi(\vec{k},\infty)|  =0~;~ |C_{\phi\phi}(\vk,\vq;\infty)| =0 ~;~ |C_{\phi\chi\chi}(\vk,\vq,\vp;\infty)|\neq 0$ with
\be \sum_{\vq}\sum_{\vp}|C_{\phi\chi\chi}(\vk,\vq,\vp;\infty)|^2   = 1 \,.\label{saturation}\ee This physical picture describes  probability or \emph{population flow}, namely that the probabilities of the various states evolve  in time in such a way that the $\langle \Psi(t)|\Psi(t)\rangle$ is constant but population and probability flows among  multiparticle states and at asymptotically long times only the stable final particle states feature non-vanishing amplitudes.

 The main goal of this section is to understand   how unitarity is manifest in the time evolution of the probabilities. At first this notion seems puzzling: at $t=0$ the initial state  has unit probability   and the intermediate and final states vanishing probability. The matrix elements connecting the initial, intermediate and final states are all perturbatively small in the couplings, yet at asymptotically late time when the initial state has decayed, the total   probability of the final state, related to the initial state by perturbative matrix elements must be unity, highlighting the non-perturbative aspects of the dynamics.

 We now study this process in detail to analyze the various time scales associated with this ``population flow''. Let us introduce
\be \mathcal{E}= \epsfi -E^\pi_k ~~;~~ \Delta\Gamma = \Gamma_\pi-\Gamma_{\phi_1}\,, \label{defs1}\ee in terms of which
\be |C_{\phi\phi}(\vk,\vq;t)|^2 = |\Mpi|^2\, e^{-\Gamma_{\phi_1}\,t}~\frac{\Big|e^{i\mathcal{E}t}~e^{-\frac{\Delta\Gamma}{2}\,t}-1 \Big|^2}{\Big[\mathcal{E}^2+ \Big( \frac{\Delta\Gamma}{2}\Big)^2\Big]}\,. \label{cfisq}\ee In the narrow width limit this expression becomes proportional to $\delta(\mathcal{E})/|\Delta\Gamma|$, in order  to find the proportionality factor we integrate  (\ref{cfisq}) in $-\infty \leq \mathcal{E}\leq \infty$ and obtain to leading order in the narrow width limit
\be |C_{\phi\phi}(\vk,\vq;t)|^2 = 2\pi ~   {|\Mpi|^2} \, \frac{\Big(e^{-\Gamma_{\phi_1}\,t}-e^{-\Gamma_\pi\,t} \Big)}{\Gamma_\pi-\Gamma_{\phi_1}} ~\delta(E^\pi_k-E^{\phi_1}_{\vq}-E^{\phi_2}_{\vk-\vq})\,. \label{Cfifina}\ee

This result is obviously non-perturbative: while in the numerator is $|\Mpi|^2 \propto g^2_{\pi}$ the denominator is \emph{also} $\propto g^2_{\pi},g^2_{\phi}$ exhibiting the non-perturbative nature of the result. At very early times $t \ll 1/\Gamma_\pi;1/\Gamma_{\phi_1}$ the total  number of $\phi_1 $ particles grows linearly with time with a rate
\be \Gamma(\pi\rightarrow \phi_1\phi_2) = 2\pi \sum_{\vq} \,  {|\Mpi|^2} \delta(E^\pi_k-E^{\phi_1}_{\vq}-E^{\phi_2}_{\vk-\vq}) \label{rate}\ee however it reaches a maximum at $t=t^*$ given by
\be t^* = \frac{\ln\Big[\frac{\Gamma_\pi}{\Gamma_{\phi_1}}\Big]}{\Gamma_\pi-\Gamma_{\phi_1}} \label{tmax}\ee and falls off exponentially on a time scale determined by the smaller of $\Gamma_\pi,\Gamma_{\phi_1}$.

\begin{figure}[h!]
\begin{center}
\includegraphics[height=3.5in,width=3in,keepaspectratio=true]{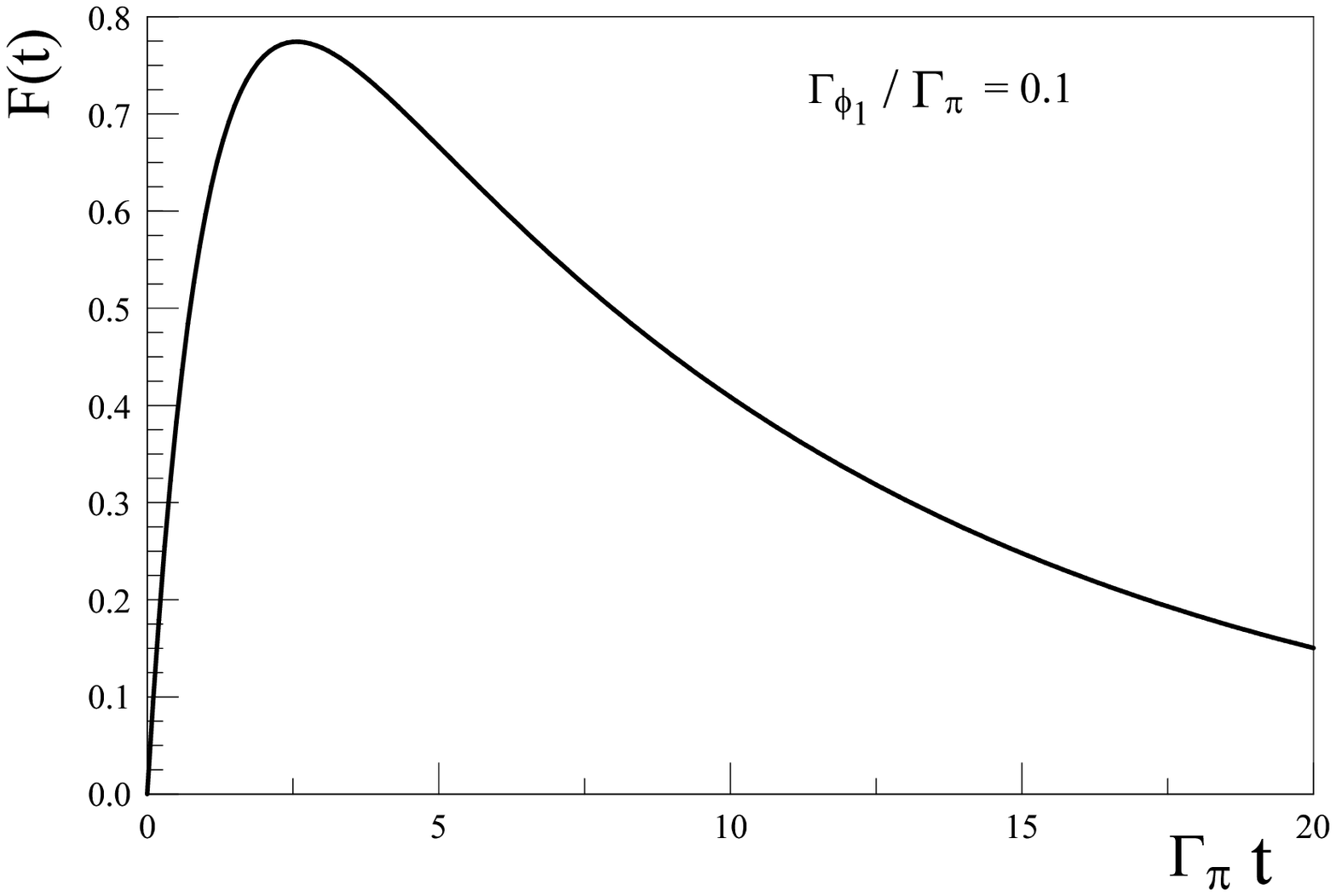}
\includegraphics[height=3.5in,width=3in,keepaspectratio=true]{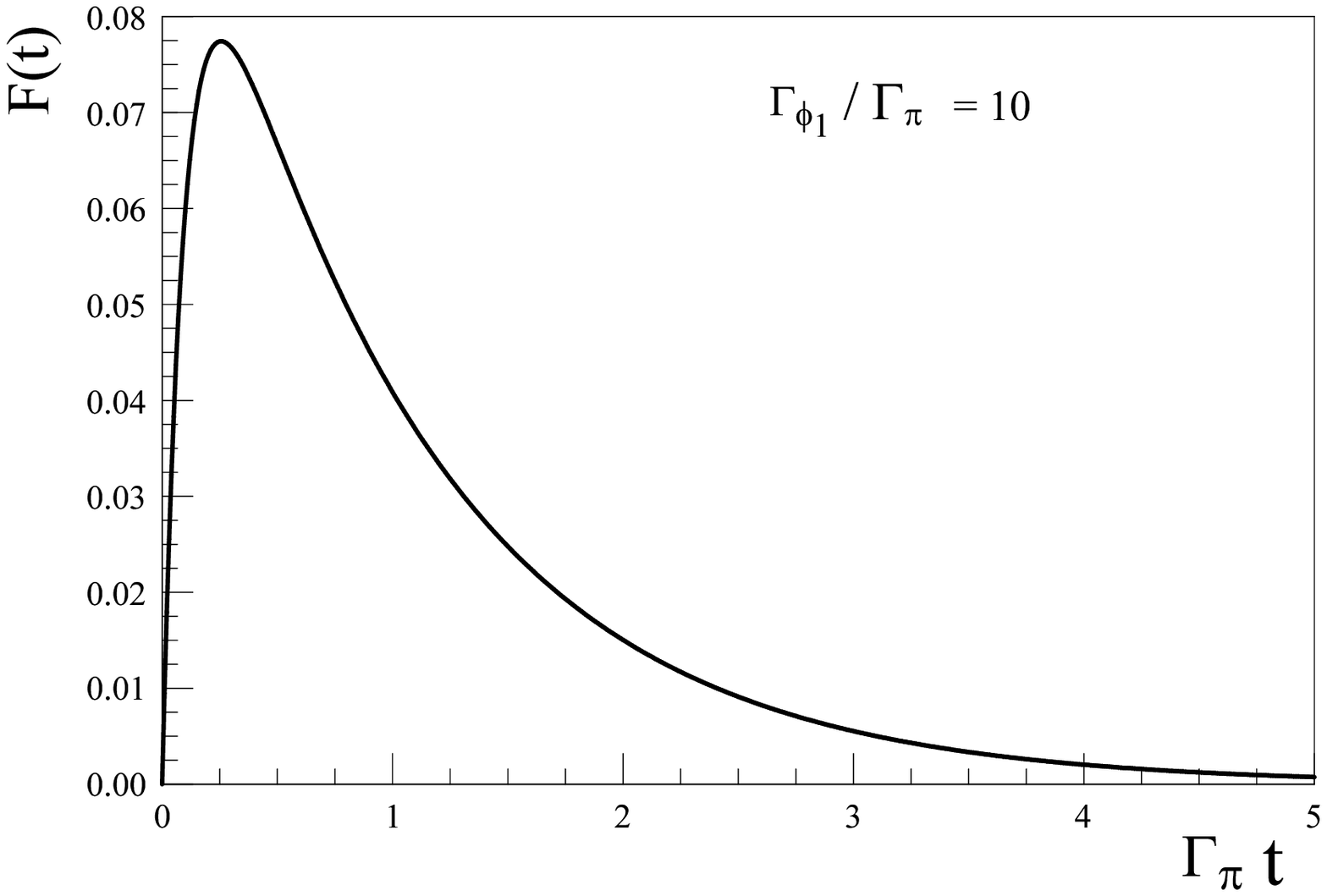}
\caption {The function $F(t) = \Big(e^{-\Gamma_{\phi_1}\,t}-e^{-\Gamma_\pi\,t} \Big)/(1-\Gamma_{\phi_1}/\Gamma_\pi)$ for $\Gamma_{\phi_1}/\Gamma_{\pi} =0.1,10$.  }
\label{fig:foft}
\end{center}
\end{figure}

In particular if the $\pi$ particle decays at rest the total number of $\phi_1$ particles is given by
\be N_{\phi_1}(t)= \sum_{\vq}|C_{\phi\phi}(\vk,\vq;t)|^2 = \frac{\Gamma(\pi\rightarrow \phi_1\phi_2)}{\Gamma_\pi} \,\Bigg[\frac{e^{-\Gamma^*_{\phi_1}\,t}-e^{-\Gamma_\pi\,t} }{{1-\frac{\Gamma^*_{\phi_1}}{\Gamma_\pi}} }\Bigg]\label{Noffi1}\ee where
\be \Gamma^*_{\phi_1}=\Gamma_{\phi_1}(q*)~~;~~q*= \frac{1}{2m_\pi}\Big[m^4_\pi+m^4_{\phi_1}+m^4_{\phi_2}-2m^2_\pi\,m^2_{\phi_1}-2m^2_\pi\,m^2_{\phi_2}-
2m^2_{\phi_1}\,m^2_{\phi_2}  \Big]^{\frac{1}{2}} \,.\label{gamafi1star}\ee
The time dependent function in the bracket in (\ref{Noffi1}) that determines the population of the resonant particles  is shown in fig. (\ref{fig:foft}) as a function of $\Gamma_\pi\,t$ for $\Gamma_{\phi_1}/\Gamma_\pi =0.1,10$, displaying the behavior discussed above: an early linear growth from the production of resonant states from the decay of the parent particle for $t \ll \Gamma_\pi,\Gamma_{\phi_1}$ growing to a maximum and falling off exponentially on the longer time scale determined by the smaller of the decay widths.

We now turn to $|C_{\phi\chi\chi}(\vk,\vq,\vp;t)|^2$ focusing first on the asymptotic $t\rightarrow \infty$ limit,   we find
\be \sum_{\vq,\vp} |C_{\phi\chi\chi}(\vk,\vq,\vp;\infty)|^2 =  \sum_{\vq,\vp}\frac{|\Mpi|^2~\, |\Mfi|^2}{\Big[ \Big(\epsfichi-E^\pi_k \Big)^2 +\Big(\frac{\Gamma_\pi}{2} \Big)^2\Big]\Big[ \Big(\epsfichi-\epsfi\Big)^2 +\Big(\frac{\Gamma_{\phi_1}}{2} \Big)^2\Big]}\,. \label{cinfi}\ee
Remarkably this expression is very similar to the asymptotic limit of the probability of a two photon state in a two-level radiative cascade in quantum optics\cite{qobooks1,mutu}.

In the narrow width limit this expression is dominated by the resonances and as usual we approximate
\be \frac{1}{\Big[ \Big(\epsfichi-E^\pi_k \Big)^2 +\Big(\frac{\Gamma_\pi}{2} \Big)^2\Big]} \rightarrow \frac{2\pi}{\Gamma_\pi}\,\delta\Big(\epsfichi-E^\pi_k \Big) \label{apro}\ee therefore
\be \sum_{\vq,\vp} |C_{\phi\chi\chi}(\vk,\vq,\vp;\infty)|^2 =  \frac{2\pi}{\Gamma_\pi}\, \sum_{\vq,\vp}\frac{|\Mpi|^2~\, |\Mfi|^2~\delta\Big(\epsfichi-E^\pi_k \Big)}{\Big[ \Big(E^\pi_k-\epsfi\Big)^2 +\Big(\frac{\Gamma_{\phi_1}}{2} \Big)^2\Big]} \label{asido}\ee where in the denominator we used the delta function to set $\epsfichi = E^\pi_k$.

Upon comparing this result with the result for $\Gamma_\pi$ given by (\ref{gamapires}) (see the definitions (\ref{epsfi},\ref{epsfichi}) we find
\be \sum_{\vq,\vp} |C_{\phi\chi\chi}(\vk,\vq,\vp;\infty)|^2 = 1\label{unitari}\ee which is the manifestation of unitarity in the asymptotic limit.

The analysis for finite time is lengthy and relegated to appendix (\ref{app:fintime}), here we present the main results.

In the narrow width limit we replace
\be \frac{1}{\Big[ \Big(\epsfi-E^\pi_k \Big)^2 +\Big(\frac{\Gamma_{\phi_1}}{2} \Big)^2\Big]} \rightarrow \frac{2\pi}{\Gamma_{\phi_1}}\,\delta\Big(\epsfi-E^\pi_k \Big) \label{apropo}\ee and to leading order in the widths it follows that
\be \sum_{\vq,\vp} |C_{\phi\chi\chi}(\vk,\vq,\vp;\infty)|^2 = (2\pi)^2 \sum_{\vq,\vp}\frac{|\Mpi|^2~\, |\Mfi|^2}{\Gamma_\pi~\Gamma_{\phi_1}}~\delta\Big(\epsfichi-E^\pi_k \Big)~\delta\Big(\epsfi-E^\pi_k \Big) \,, \label{cfinity}\ee
combining this expression with (\ref{dcs}) and (\ref{totalsum}) we finally find
\bea \sum_{\vq,\vp} |C_{\phi\chi\chi}(\vk,\vq,\vp;t)|^2   = &&(2\pi)^2 \sum_{\vq,\vp}\frac{|\Mpi|^2~\, |\Mfi|^2}{\Gamma_\pi~\Gamma_{\phi_1}~(\Gamma_\pi-\Gamma_{\phi_1})} ~\Bigg\{\Gamma_\pi~(1-e^{- {\Gamma_{\phi_1}} \,t})-\Gamma_{\phi_1}~(1-e^{- {\Gamma_{\pi}} \,t})  \Bigg\}   \nonumber   \\ && \times \delta\Big(\epsfichi-E^\pi_k \Big)~\delta\Big(\epsfi-E^\pi_k \Big) \,  .\label{coftifi}\eea This expression may be simplified by realizing that $\Gamma_{\phi_1}(\vq)$ depends on $\vq$ but not on $\vp$ and  because of the second delta function in (\ref{coftifi}) we can set $\delta\Big(\epsfichi-E^\pi_k \Big) \rightarrow \delta\Big(\epsfichi-\epsfi\Big)= \delta(E^{\phi_1}_{\vq}-E^{\chi_1}_{\vp}-E^{\chi_2}_{\vq-\vp} )$ in the first delta function, using the result (\ref{Gammafi1}) we finally find
\be  \sum_{\vq,\vp} |C_{\phi\chi\chi}(\vk,\vq,\vp;t)|^2   =   \frac{2\pi}{\Gamma_\pi }   \sum_{\vq} |\Mpi|^2 \Bigg[\frac{\Gamma_\pi~(1-e^{- {\Gamma_{\phi_1}} \,t})-\Gamma_{\phi_1}~(1-e^{- {\Gamma_{\pi}} \,t})}{(\Gamma_\pi-\Gamma_{\phi_1})}  \Bigg]   \delta\Big(E^\pi_k- E^{\phi_2}_{\vk-\vq}- E^{\phi_1}_{\vq} \Big) \,  .\label{coftifi2}\ee Combining this result with (\ref{Cfifina}) and using (\ref{gamapifini}) we find
\be \sum_{\vq}|C_{\phi\phi}(\vk,\vq;t)|^2+ \sum_{\vq,\vp} |C_{\phi\chi\chi}(\vk,\vq,\vp;t)|^2 = 1-e^{- {\Gamma_{\pi}} \,t} = 1- |C_\pi(t)|^2 \,, \label{totalsumita} \ee this is the final result confirming unitary time evolution and the flow of population from the initial through the intermediate to the final state.

Therefore an important conclusion that follows from this analysis is that the amplitudes (\ref{Cfifit},\ref{Cfichi2}) are the correct ones insofar as they manifestly satisfy unitarity.

\begin{figure}[h!]
\begin{center}
\includegraphics[height=3.5in,width=3.2in,keepaspectratio=true]{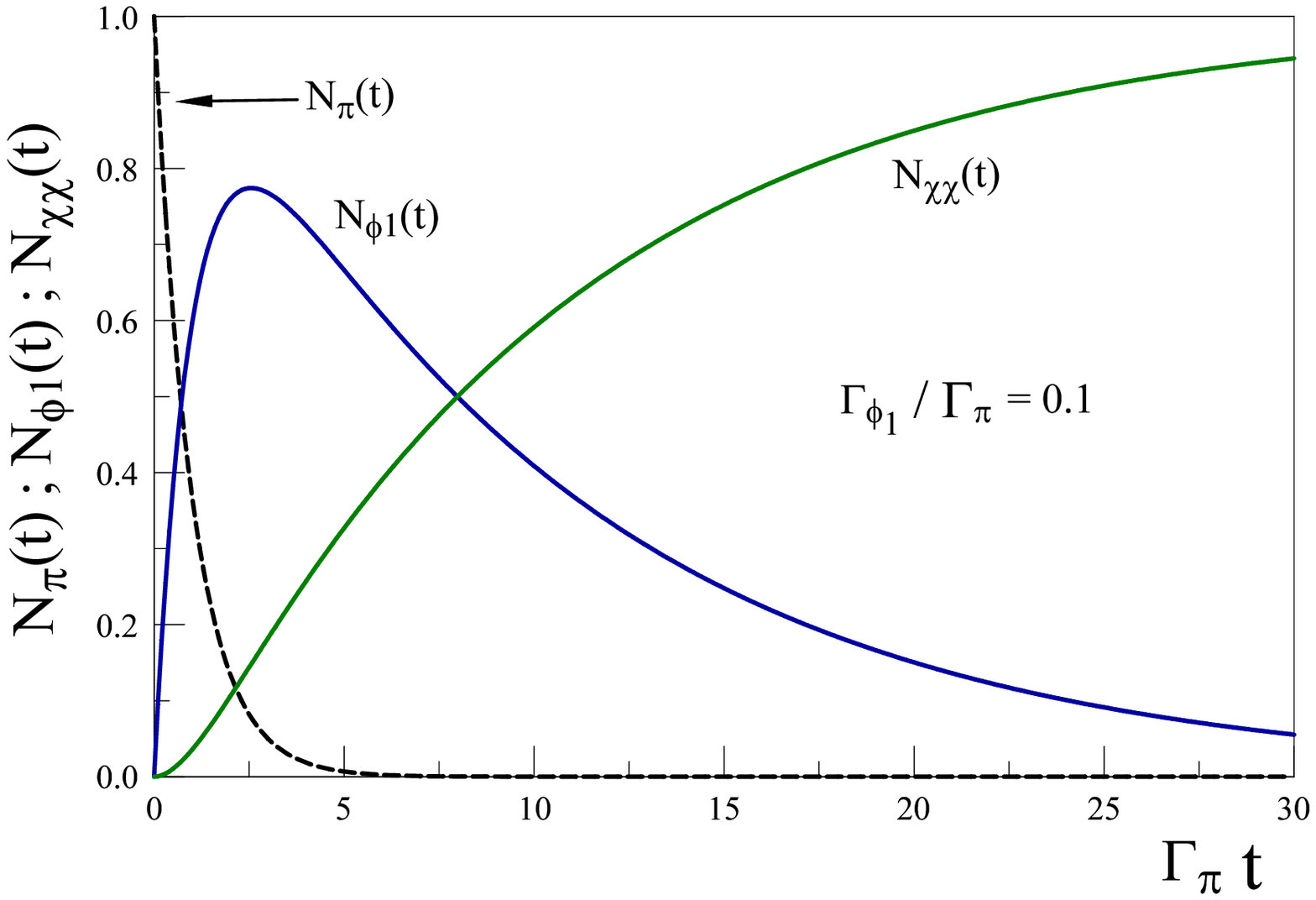}
\includegraphics[height=3.5in,width=3.2in,keepaspectratio=true]{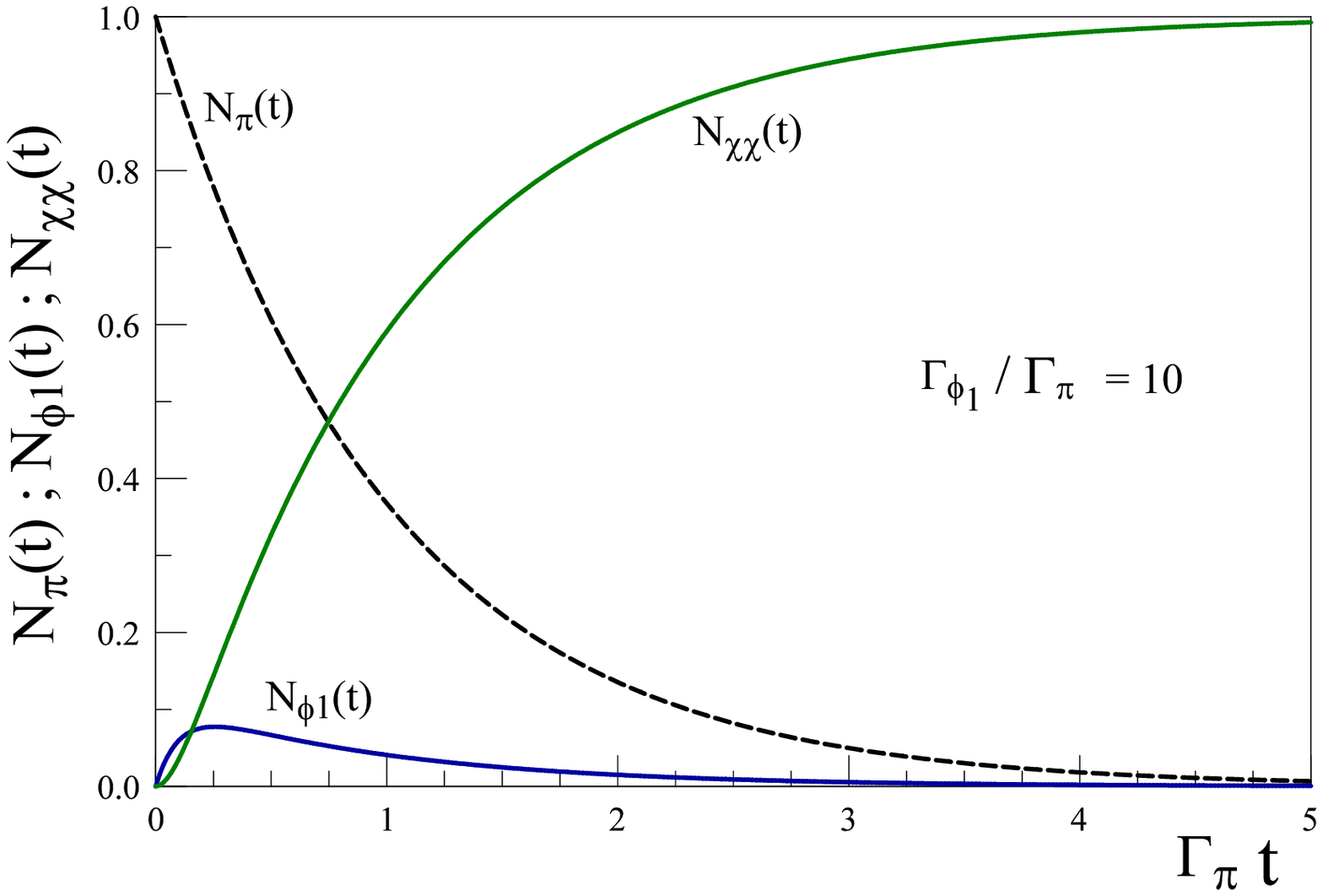}
\caption {The number of $\pi$-particles $N_\pi(t)$, resonant states $N_{\phi_1}(t)$ and $\chi$-pairs $N_{\chi\chi}(t)$ vs. $\Gamma_\pi\,t$   for $\Gamma_{\phi_1}/\Gamma_{\pi} =0.1,10$.  }
\label{fig:numbers}
\end{center}
\end{figure}

In particular when the $\pi$ particle decays at rest, from the result (\ref{coftifi2}) and the identification (\ref{pairchi}) the total number of \emph{pairs} of $\chi_1,\chi_2$ particles is given by
\be N_{\chi\chi}(t)=\sum_{\vq,\vp} |C_{\phi\chi\chi}(\vk,\vq,\vp;t)|^2  = \Bigg[\frac{\Gamma_\pi~(1-e^{- {\Gamma^*_{\phi_1} } \,t})-\Gamma^*_{\phi_1} ~(1-e^{- {\Gamma_{\pi}} \,t})}{(\Gamma_\pi-\Gamma^*_{\phi_1} )}  \Bigg] \label{numpair}\ee where $\Gamma^*_{\phi_1} $ is given by (\ref{gamafi1star}). In summary:
\bea N_\pi(t) & = & e^{-\Gamma_\pi\,t} \label{npi}\\
N_{\phi_1}(t) & = & \frac{e^{-\Gamma^*_{\phi_1}\,t}-e^{-\Gamma_\pi\,t} }{{1-\frac{\Gamma^*_{\phi_1}}{\Gamma_\pi}} } \label{nfi1}\\
N_{\chi\chi}(t) & = & \frac{\Gamma_\pi~(1-e^{- {\Gamma^*_{\phi_1} } \,t})-\Gamma^*_{\phi_1} ~(1-e^{- {\Gamma_{\pi}} \,t})}{(\Gamma_\pi-\Gamma^*_{\phi_1} )}\,, \label{nchichi}\eea with
\be  N_\pi(t)+N_{\phi_1}(t)+N_{\chi\chi}(t)=1\,. \label{sumaeq1}\ee The populations are shown in fig. (\ref{fig:numbers}) as a function of time for $\Gamma_{\phi_1}/\Gamma_{\pi} =0.1,10$.

\vspace{3mm}

\textbf{Discussion:}
The expressions (\ref{npi}-\ref{nchichi}) and the figures (\ref{fig:foft},\ref{fig:numbers}) display the   behavior of the populations of resonant and final states and describe the main time dependent  physical phenomena of the cascade decay. As discussed above at early time $t \ll 1/\Gamma_\pi,1/\Gamma_{\phi_1}$ the number of resonances grows linearly in t as $N_{\phi_1}(t) \simeq \Gamma_\pi t$, as the decay the the parent particle increases the population of resonances, riches a maximum at $t^*$ given by (\ref{tmax}) and decays on the longer time scale. However the population of the final state $\chi\chi\phi_2$ at early time grows as $ N_{\chi\chi}(t) \simeq \Gamma_\pi\,\Gamma_{\phi}t^2/2$, namely much \emph{slower} than the population of the intermediate resonant state. This is a result of a second order process as the build-up of the final state requires first to build up the population of the resonant state from the decay of the parent state as $\simeq \Gamma_\pi\,t$ and the build-up of the final state from the decay of the \emph{populated} intermediate resonant state as $\Gamma_{\phi_1}\,t$. The behavior of the populations at intermediate and long times depends on the particular cases   $\Gamma_\pi \gg \Gamma_{\phi_1}$ and $\Gamma_\pi \ll \Gamma_{\phi_1}$.

\vspace{1mm}

\begin{itemize}
\item{$\mathbf{\Gamma_\pi \gg \Gamma_{\phi_1}}$: In this case the slowly decaying intermediate resonant state acts as a ``bottleneck'', the rapid decay of the parent particle   builds up the population of the resonant states which grows   to a maximum at $t^*$ with an amplitude $N_{\phi_1}(t^*)$. For $\Gamma_\pi \gg \Gamma_{\phi_1}$ it follows that $e^{-\Gamma_\pi\,t^*} \ll 1~~;~~e^{-\Gamma_{\phi_1}\,t^* } \sim 1$ and  $N_{\phi_1}(t^*)\sim 1 $. The decay of the resonant state into the final state particles occurs on a \emph{much slower} time scale of the order of $1/\Gamma_{\phi_1}$. In this limit the cascade decay can be described \emph{sequentially} as $\pi \rightarrow \phi_2\phi_1 ~;~ \phi_2\phi_1\rightarrow \chi_1\chi_2\phi_2$ where the intermediate resonant state attains an amplitude $\simeq 1$, nearly saturating unitarity  on a short time scale  $t\simeq t^* = \ln[ {\Gamma_\pi}/{\Gamma_{\phi_1}} ]/(\Gamma_\pi-\Gamma_{\phi_1}) $ and decays slowly on the time scale $1/\Gamma_{\phi_1}$. For $t > 1/\Gamma_\pi$ it follows that the population of the final state $N_{\chi\chi}(t) \simeq (1-e^{- {\Gamma^*_{\phi_1} } \,t})$ consistently with the decay of an initial resonant state of nearly unit amplitude.   }

\item{$\mathbf{\Gamma_\pi \ll \Gamma_{\phi_1}}$: In this case   the intermediate resonant state decays on time scales shorter than that of the parent particle, as a result there is very little population build-up of the resonant state and the initial population of the parent particle is ``transferred'' directly to the final state on a time scale $\simeq 1/\Gamma_{\pi}$. In this limit $\Gamma_{\phi_1}\,t^* \gg 1$ and $N_{\phi_1}(t^*) \simeq  e^{-\Gamma_\pi\,t^*}~\Gamma_\pi/\Gamma_{\phi_1} \ll 1$, and for $t \gtrsim t^*$ it follows that $N_{\chi\chi}(t) \simeq (1-e^{- {\Gamma_{\pi} } \,t})$ describing the build up of the population of the final states directly from the decay of the parent particle. In this limit the decay of the parent particle can be described as a direct decay into the final states as the intermediate resonant state is so short-lived that the population of resonances does not build up substantially. }

    \item{$\mathbf{\Gamma_\pi = \Gamma_{\phi_1}}$: in this (unlikely) case it is straightforward to find
     \be N_{\phi_1}(t) = \Gamma_\pi\,t~ e^{-\Gamma_\pi\,t} ~~;~~  N_{\chi\chi}(t) = 1-e^{-\Gamma_\pi\,t}\Big[1+\Gamma_\pi\,t\Big] ~~;~~t^*= 1/\Gamma_\pi \,. \label{eqgams}\ee
    }

\end{itemize}

%%% new addition # 1 &&&&

The ``bottleneck'' in the case when $\mathbf{\Gamma_\pi \gg \Gamma_{\phi_1}}$ is reminiscent of a similar phenomenon in the production of Helium during Big Bang Nucleosynthesis (BBN) via the formation of a deuteron bound state in $n-p$ collisions. The deuteron is photo-dissociated by the high energy photons in the blackbody tail and does not form until the ambient temperature falls to about $80 \, \mathrm{keV}$ resulting in a delayed transition from the initial $n-p$ to the final state. However, although there is a similarity in that the decay to the final state is delayed, the physical reasons are different: in the case under consideration in this article, the delayed decay to the final state is a consequence of the on-shell formation of a long-lived resonant state, whereas in BBN the delay is a consequence of photodissociation of the deuteron intermediate state which does not form until the temperature falls well below  the binding energy of the bound state $\approx 2 \,\mathrm{MeV}$. Once the deuteron is formed the fusion reactions end up in $^{4}He$ very fast.

%% end of new addition #1 %%
\vspace{2mm}

\textbf{Generality of the results:}

Although we focused on the specific example given by the interaction Hamiltonian (\ref{HI}) the procedure leading to the equations for the amplitudes is general. In particular we have left the matrix elements (\ref{Mpi},\ref{Mfi}) indicated without using their explicit form and the final results for the amplitudes depend solely on the transition matrix elements, decay widths of the parent and resonant states and single particle energies. Therefore the extension of the above results to any other theory in which a cascade decay proceeds via an intermediate resonant state is straightforward and can be described by the above results by replacing the proper matrix elements and decay widths.

\vspace{3mm}

\section{Field theoretic generalization of Wigner-Weisskopf:}

The Wigner-Weisskopf theory of spontaneous emission\cite{ww}  plays an important role in quantum optics\cite{qobooks1,qobooks2,qobooks3} and in particle physics it is one of the main approaches to study the dynamics of the $K^0-\overline{K^0}$ system\cite{wwkaon,wwcp} (also $B^0-\overline{B^0}$). A quantum field theoretical generalization of this important method was provided in ref.\cite{wwnos} with an extension in cosmology in refs.\cite{wwds,lellods}. However, to the best of our knowledge the method has not been extended to the case of radiative cascades. This is the goal of this section.

 Again, writing the Hamiltonian   as   $H=H_0+H_I$, and passing to the interaction picture  instead of using the Schrodinger picture, as in (\ref{timeevol}), the time evolution of the quantum state is given by

\be \label{inttimeevol}
i \frac{d}{dt} |\Psi(t)\rangle_I = H_I(t)|\Psi(t)\rangle_I
\ee where $H_I(t)$ is the interaction Hamiltonian in the interaction picture.

The  state $|\Psi(t)\rangle_I$ is then expanded in the basis of free particle Fock states $|m\rangle$ eigenstates of $H_0$, namely
\be |\Psi(t)\rangle_I = \sum_m A_m(t) |m\rangle \,. \label{ip}\ee
Using the orthogonality relation on any state expanded in this fashion  leads to the following relation

\be
\dot{A}_{n}(t) = -i\sum_m \langle n |H_I(t)| m \rangle A_{m}(t) \,.\label{ipeq}
\ee

This allows for a \emph{time dependent} interaction Hamiltonian, a situation common to field theories in curved spacetime\cite{wwds,lellods}. Specifically, the method described below, when generalized to the case of an expanding cosmology is amenable to be implemented to study the cascade decay of inflationary quanta discussed in ref.\cite{lellods}. We intend to apply the method developed here to the case of the  cosmological cascade decay  in future studies.

 For the purposes of this work, ensuring that the Wigner-Weisskopf procedure reproduces the results obtained via Laplace transform will confirm that this real time method provides a   non-perturbative resummation that yields the correct time evolution at least in the cases where it can be compared to known results.

 In the interaction picture the quantum state that describes the cascade decay is given by

\be \label{intstate}
|\Psi(t)\rangle_I =  A_\pi(\vec{k},t)\big|\pi_{\vec{k}}\rangle + \sum_{\vq}\,A_{\phi\phi}(\vk,\vq;t)\,\big|\phi_{1,\vq}\phi_{2,\vk-\vq} \rangle + \sum_{\vq;\vp} A_{\phi\chi\chi}(\vk,\vq,\vp;t)\,\big|\phi_{2,\vk-\vq}\chi_{1,\vp}\chi_{2,\vq-\vp} \rangle  +\cdots
\ee   When the interaction Hamiltonian in the Schroedinger picture is time independent as is the case in Minkowski space time in absence of external sources,    the coefficients in this expression differ from those in Section \ref{sec:model} by the relation $C = A e^{-iE t}$, where $E$ is the energy (eigenvalue of $H_0$) of the particular state in the expansion, this is the difference between the two pictures.

Restricting attention to Minkowski spacetime in absence of explicit time dependent sources, the orthogonality relations lead to matrix elements of the interaction Hamiltonian given by

\be
\mathcal{M}_{mn} (t) = \langle m | H_I(t) | n \rangle = e^{i(E_m-E_n)t} \langle m|H_I|n\rangle ~~;~~ H_I(t)= e^{iH_0t}\,H_I\,e^{-iH_0t}.
\ee
 Again, consider the interaction Hamiltonian given by (\ref{HI}) and the initial conditions given by (\ref{inistate}). To simplify notation, the definitions introduced in Eqs (\ref{Mpi}-\ref{epsfichi}) will be used leading to the following equations for the coefficients

\be \label{api}
\dot{A}_{\pi}(\vk,t) = -i\sum_{\vq} \Mpi A_{\phi \phi}(\vk,\vq,t)\, e^{i(E_k^{\pi}-\epsfi)t}~~;~~A_\pi(\vk,0)=1\,, \ee

\be \label{aphiphi}
\begin{split}
 \dot{A}_{\phi \phi}(\vk,\vq,t) =  &-i \Mpi^{*} A_{\pi}(\vk,t)\, e^{i(\epsfi-E^{\pi}_k)t} \\
& -i \sum_{\vp} \Mfi A_{\phi \chi \chi}(\vk,\vq,\vp,t)\, e^{i(\epsfi - \epsfichi)t}~~;~~ {A}_{\phi \phi}(\vk,\vp,0)=0 \,,
\end{split}
\ee

\be \label{aphichichi}
\dot{A}_{\phi \chi \chi}(\vk,\vq,\vp,t) = -i \Mfi^{*} e^{-i(\epsfi - \epsfichi)t} A_{\phi \phi}(\vk,\vq,t)~~;~~ {A}_{\phi \chi \chi}(\vk,\vq,\vp,0)=0\,.
\ee

Again we solve the hierarchy from the bottom up. The solution of (\ref{aphichichi}) is
\be {A}_{\phi \chi \chi}(\vk,\vq,\vp,t) =  -i \Mfi^{*} ~\int^t_0  e^{-i(\epsfi - \epsfichi)t'} A_{\phi \phi}(\vk,\vq,t')dt' \, \label{aphichichisol} \ee and combining (\ref{aphiphi}) and (\ref{aphichichisol}) yields (now suppressing the momenta in the arguments)

\be
\dot{A}_{\phi \phi}(t) + \sum_{\vp} |\Mfi|^2 \int^{t}_{0} dt' e^{-i (\epsfi - \epsfichi)(t'-t)} A_{\phi \phi}(t') = -i \Mpi^{*} e^{-i (E_{\pi}-\epsfi) t} A_{\pi}(t)\,.\label{intedife}
\ee

A perturbative solution to this integro-differential equation in terms of $A_\pi$ is straightforward, however, it leads to resonant denominators and its eventual breakdown. Instead we implement a non-perturbative approach that provides a resummation that  incorporates consistently the width of the resonant state.

 In order to implement this method to solve (\ref{intedife}), let us first focus on the homogeneous case neglecting the right hand side. Consider

\be \label{homo}
\dot{A}^{H}_{\phi \phi}(t) + \sum_{\vp} |\Mfi|^2 \int^{t}_{0} dt' e^{-i (\epsfi - \epsfichi)(t'-t)} A^{H}_{\phi \phi}(t') = 0 ~~;~~ {A}^{H}_{\phi \phi}(0) =1 \,.
\ee

This equation simplifies by implementing a Markovian approximation which is justified in weak coupling. Since  the term inside the integrand   $|\Mfi|^2 \sim g_{\phi}^2 \ll 1$ it follows that $\dot{A}^{H}_{\phi \phi}\propto g_{\phi}^2 \ll 1$, namely the amplitudes vary \emph{slowly} in time. To clearly see the nature of the  approximation, the following is introduced

\be \label{wzero}
W^{\phi}_0(t,t') = \sum_{\vp} |\Mfi|^2 \int^{t'}_{0} dt'' e^{-i(\epsfi - \epsfichi)(t''-t)}
\ee which has the properties

\be
\frac{d}{dt'} W^{\phi}_0(t,t') =  \sum_{\vp} |\Mfi|^2 e^{-i(\epsfi - \epsfichi)(t'-t)}  \sim \mathcal{O}(g_{\phi}^2)~~;~~ W^{\phi}_0(t,0) = 0
\ee An integration by parts produces

\be
\int^t_0 dt' \frac{d}{dt'} W^{\phi}_0(t,t') A_{\phi \phi}(t') = W^{\phi}_0(t,t) A_{\phi \phi} (t) - \int_0^t dt' \dot{A}_{\phi \phi}(t') W^{\phi}_0(t,t')
\ee This can be repeated systematically, producing higher order derivatives by using the natural definition

\be
W^{\phi}_N(t,t') = \int^{t'}_0 dt'' W^{\phi}_{N-1}(t,t'') ~~;~~ W^{\phi}_N(t,0) = 0 \,.
\ee Repeated integration by parts produces the series

\be
\int^t_0 dt' \frac{d}{dt'} W^{\phi}_0(t,t') A_{\phi \phi}(t') = W^{\phi}_0(t,t) A_{\phi \phi} (t) -  W^{\phi}_1(t,t) \dot{A}_{\phi \phi} (t) + W^{\phi}_2(t,t)
\ddot{A}_{\phi \phi} (t) + ...
\ee where each term has a multiplicative factor $W^{\phi}_N \sim g_{\phi}^2$ and $\dot{A} \propto g_{\phi}^2;\ddot{A} \propto g_{\phi}^4$ etc. Truncating this series to leading order, namely keeping only $W^{\phi}_0$ gives

\be
\dot{A}^{H}_{\phi \phi} + W^{\phi}_0(t,t)  A^{H}_{\phi \phi}(t) = 0  ~~;~~ {A}^{H}_{\phi \phi}(0) =1 \,. \label{marko}
\ee This makes it apparent that the lowest order solution results from keeping only $W_0$, this is the Markovian approximation. So to lowest order, the homogenous solution is written

\be
A^{H}_{\phi \phi}(t) = e^{-\int^t_0 dt' W^{\phi}_0(t',t')}
\ee where

\bea \label{homocoeff}
\int^t_0 dt' W^{\phi}_0(t',t') & = &  i t ~\sum_{\vp} \frac{|\Mfi|^2}{(\epsfi - \epsfichi)} \Bigg\{\Bigg(1-\frac{\sin(\epsfi-\epsfichi)\,t}{(\epsfi-\epsfichi)\,t}\Bigg) \nonumber \\ & -i & \Bigg(\frac{1- \cos(\epsfi-\epsfichi)\,t}{(\epsfi-\epsfichi)t}\Bigg)\Bigg\}
\eea

 We prove in appendix (\ref{appB}) (see also ref.\cite{wwds}) that in  the long time limit for time scales that are much larger than the energy uncertainty ($t \gg 1/(\epsfi-\epsfichi)$),
\be
\int^t_0 dt' W^{\phi}_0(t',t') \rightarrow t \left(i \mathcal{P} \sum_{\vp} \frac{|\Mfi|^2}{\epsfi-\epsfichi} + \pi \sum_{\vq} |\Mfi(p,q)|^2 \delta (\epsfi - \epsfichi)  \right)
\ee However, the same result can be obtained from replacing $\epsfi -\epsfichi \rightarrow \epsfi - \epsfichi + i \varepsilon$ with $\varepsilon \rightarrow 0^+$ and $t=t'\rightarrow \infty$ in (\ref{wzero}) which yields

\be \lim_{t  \rightarrow \infty}\, \int^t_0 dt'   W_0(t',t') = i\,t\, \sum_{\vp} \frac{|\Mfi|^2}{(\epsfi - \epsfichi + i \varepsilon )}  = \Bigg(i\Delta E^{\phi_1}+ \frac{\Gamma_{\phi_1}}{2} \Bigg)\,t \label{longtime}
\ee  where  $\Delta E^{\phi_1}$ and $\Gamma_{\phi_1}$ are given by (\ref{deltaEfi1},\ref{Gammafi1}) respectively, and  the homogenous solution becomes

\be
A^{H}_{\phi \phi}(t) =   e^{-i\,\Delta E^{\phi_1}\,t}~e^{-\frac{\Gamma_{\phi_1}}{2}\, t}\,.
\ee

Now with the left hand side of eqn. (\ref{intedife}) replaced by the Markovian approximation (\ref{marko}) and using  the above result,  the full solution of (\ref{intedife}) is given by

\be
A_{\phi \phi}(t) = -i \Mpi^{*} e^{-i(\Delta E_{\phi_1} - i\frac{\Gamma_{\phi_1}}{2} )\,t} \int_0^t dt' \,A_{\pi}(t')\, e^{-i(E^{\pi} - \epsfi - \Delta E_{\phi_1} + i\frac{\Gamma_{\phi_1}}{2} )\,t'}\,.\label{solumarki}
\ee Inserting this solution  into Eq (\ref{api}) we obtain

\be
\dot{A}_{\pi}(t) + \sum_{\vq} |\Mpi|^2\int_0^t dt' e^{i(E^{\pi} - \epsfi -\Delta E_{\phi_1} + i\frac{\Gamma_{\phi_1}}{2})(t-t')} A_{\pi}(t') = 0 ~~;~~ {A}_{\pi}(0) =1 \,.
\ee

At this stage, we implement again a  Markovian approximation as described above,  which is justified in this case by the weak coupling   $g_\pi \ll 1$ so that the integrand can be treated as slowly varying and carrying out the same expansion as above and keeping the lowest order we obtain

\be
\dot{A}_{\pi}(t) + W_0^{\pi}(t,t) A_{\pi}(t) = 0 ~~;~~ {A}_{\pi}(0) =1 \,,
\ee where

\be
W_0^{\pi}(t,t') = \sum_{\vq} |\Mpi|^2 \int^{t'}_{0} dt''\, e^{-i(E^{\pi} - \epsfi -\Delta E_{\phi_1} + i\frac{\Gamma_{\phi_1}}{2})(t''-t)}
\ee and in the same manner as before the solution is now given by

\be
A_{\pi}(t) = e^{-\int^t_0 dt' W^{\pi}_0(t',t')}\label{apisolu}
\ee where again taking the long time limit

\be
  \int^{t}_0 dt' W^{\pi}_0(t',t') =i\,t  \sum_{\vq} \frac{|\Mpi|^2}{E^{\pi} - \epsfi - \Delta E_{\phi_1} + i \frac{\Gamma_{\phi_1}}{2} + i\varepsilon} \,,
\ee  where for $\Gamma_{\phi_1}\neq 0$ one can neglect $\varepsilon$. The solution now becomes

\be
 A_{\pi}(t) = e^{-i \Delta E^{\pi}\,t} e^{- \frac{\Gamma_{\pi}}{2}\,t} \label{apisolfinal}
\ee where

\be
\Delta E^{\pi} = \sum_{\vq} \frac{|\Mpi|^2\,\Big(E^{\pi} - \epsfi - \Delta E_{\phi_1}\Big)}{\Big(E^{\pi} - \epsfi - \Delta E_{\phi_1}\Big)^2 + \Big(\frac{\Gamma_{\phi_1}}{2} \Big)^2} \label{delpi2}
\ee

\be
\Gamma_{\pi}(k)  = \sum_{\vq} \frac{|\Mpi|^2 \,\Gamma_{\phi_1}}{\Big(E^{\pi} - \epsfi - \Delta E_{\phi_1}\Big)^2 + \Big(\frac{\Gamma_{\phi_1}}{2} \Big)^2}\,.\label{gamapi2}\ee

This is now   used to solve Eq (\ref{aphiphi}) resulting in the following expression

\be
\begin{split}
& A_{\phi \phi}(t)  = -i \Mpi^{*}e^{-i(\Delta E_{\phi_1} - i\frac{\Gamma_{\phi_1}}{2} )\,t} \int_0^t dt' e^{-i(E^{\pi} - \epsfi + \Delta E^{\pi} -\Delta E_{\phi_1} - \frac{i}{2}(\Gamma_{\pi}  - \Gamma_{\phi_1} ))\,t'} \\
& = \frac{ \Mpi^{*} e^{-i(\Delta E_{\phi_1} - i\frac{\Gamma_{\phi_1}}{2} )\,t}}{E^{\pi} - \epsfi + \Delta E^{\pi} -\Delta E_{\phi_1} - \frac{i}{2}(\Gamma_{\pi}  - \Gamma_{\phi_1})}~\Bigg[ e^{-i(E^{\pi} - \epsfi + \Delta E^{\pi} -\Delta E_{\phi_1} - \frac{i}{2}(\Gamma_{\pi}  - \Gamma_{\phi_1} ))t} - 1\Bigg]
\end{split}
\ee Absorbing the energy shifts  into renormalizations of the mass as before, namely

\be
 E^{\pi} = E_0^{\pi}  + \Delta E^{\pi}  ~~;~~ \epsfi =  E^0_{\phi_1}  + \Delta E_{\phi}  + E_{\phi_2}
\ee and inserting the (renormalized) solution above into (\ref{aphichichisol}) we finally find

\be
\begin{split}
& A_{\phi \chi \chi}(t) = \frac{-i \Mpi^{*} \Mfi^{*}}{E^{\pi} - \epsfi - \frac{i}{2}(\Gamma_{\pi}-\Gamma_{\phi_1})} \int^t_0 dt' \left[ e^{-i(E^{\pi} - \epsfichi - i \frac{\Gamma_{\pi}}{2})t'} - e^{-i(\epsfi - \epsfichi - i \frac{\Gamma_{\phi_1}}{2})t'} \right] \\
& = \frac{ \Mpi^{*} \Mfi^{*}}{E^{\pi} - \epsfi - \frac{i}{2}(\Gamma_{\pi}-\Gamma_{\phi_1})} \left[ \frac{ e^{-i(E^{\pi}  - \epsfichi - i \frac{\Gamma_{\pi}}{2})t} - 1}{E^{\pi} - \epsfichi - i \frac{\Gamma_{\pi}}{2}} - \frac{ e^{-i(\epsfi - \epsfichi - i \frac{\Gamma_{\phi_1}}{2})t} -1 }{\epsfi - \epsfichi- i \frac{\Gamma_{\phi_1}}{2}} \right]
\end{split}
\ee Passing back to the Schrodinger picture and after renormalization of the single particle energies it follows that
\bea && {A}_{\pi}(t)\rightarrow e^{-iE^\pi\,t}\,{A}_{\pi}(t)   =  C_\pi(t) \label{CApi}\\
&& A_{\phi \phi}(t) \rightarrow  e^{-i\epsfi\,t}\,A_{\phi \phi}(t)  =   C_{\phi\phi}(t) \label{CAfi}\\
&& A_{\phi \chi \chi}(t) \rightarrow e^{-i\epsfichi\,t}\, A_{\phi \chi \chi}(t)   =   C_{\phi\chi\chi}(t)\,, \label{CAchi}\eea  finally matching   the results   (\ref{cpioft},\ref{Cfifit}, \ref{Cfichifin}) that were obtained via Laplace transform.

 \vspace{2mm}

 \textbf{Discussion:}
 The analysis above shows that the field theoretical generalization of the Wigner-Weisskopf method provides a real time realization of the non-perturbative resummation akin to the Dyson resummation of self-energies in the propagators and yields a resummation of secular terms that grow in time. The Markovian approximation based on a derivative expansion that relies on a separation of time scales valid in the weak coupling regime in the long time limit is akin to the Breit-Wigner or narrow width approximation in that it captures reliably the decay of resonances. Furthermore, the equivalence with  the results of the previous section, clearly proves that the Wigner Weisskopf method is manifestly  unitary.

 While both methods, are equivalent in \emph{Minkowski space time}, the quantum field theoretical Wigner-Weisskopf method features the distinct advantage of direct applicability in the cosmological context wherein the expansion implies an explicitly time dependent interaction Hamiltonian \emph{in the Schroedinger picture} as a consequence of the cosmological expansion. To lowest order this method has been applied in cosmology in refs.\cite{wwds,wwnos}, however the purpose of this work is to extend it to the hitherto unexplored case of \emph{radiative cascade decay}. The results of this section indicate the reliability of the method  thereby bolstering the case for its implementation in cosmology, which will be the subject of future study.

\section{Possible phenomenological consequences and correlations.}
\subsection{Possible phenomenological consequences:}
\begin{figure}[h!]
\begin{center}
\includegraphics[height=3.5in,width=3in,keepaspectratio=true]{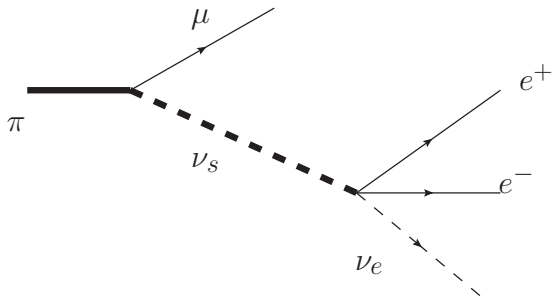}
\caption {A \emph{possible} process: $\pi \rightarrow \mu \nu_s \rightarrow e^+e^-\,\nu_e$.  }
\label{fig:appearance}
\end{center}
\end{figure}

A \emph{possible} phenomenological consequence may emerge if heavy sterile neutrinos $\nu_s$ exist and mix with active neutrinos. For example, consider the case of $\nu_\mu-\nu_s;\nu_e-\nu_s$ mixing in $\pi$-decay, if $\nu_s$ features a mass\footnote{We here simply refer to $\nu_s$  as  the sterile-like \emph{mass eigenstate}.} $2m_e < m_s < m_\pi-m_\mu$ (here we neglect the mass of the ``active'' neutrino mass eigenstates) then the intermediate state with a $\nu_s$ becomes resonant and can decay either via charged or neutral  current   interactions into $e^+e^-\nu_e$. This process is depicted in the Fermi limit in fig. (\ref{fig:appearance}) and would correspond to an \emph{appearance} contribution.

If the lifetime of the heavy sterile neutrino is very long for example of the order of the baseline in long baseline neutrino experiments, the process above will yield a contribution to the appearance probability. If on the other hand the lifetime of the putative sterile neutrino is short, then it can decay on distances shorter than the oscillation length of active neutrinos and this process would contribute to the appearance probability with oscillations and the concomitant distortion in the energy spectrum, but on length scales shorter than the oscillation length. Clearly such process will be suppressed by a product $U_{es}U_{\mu s}$. Furthermore unlike oscillations of   active neutrinos, the time dependence in this case would be damped exponentially on a scale of the order of the decay length of the sterile neutrino.
For example if the heavy sterile neutrino decays via a charged current vertex, and neglecting the electron mass
\be \Gamma_{\nu_s} \simeq \frac{G^2_F\,m^5_{\nu_s}}{192\pi^3}\,|U_{es}|^2 \label{steriledecay}\ee with a decay length
\be c\tau_{\nu_s} \simeq \frac{860}{|U_{es}|^2} \Bigg(\frac{100\,\mathrm{MeV}}{m_{\nu_s}}\Bigg)^5\,\mathrm{mts}\,.\label{decle}\ee
For $|U_{es}| \ll 1$ the maximum in the probability of the intermediate state with the sterile neutrino resonance occurs at a distance
\be c t^* \sim     -10 \ln|U_{es}|  \,\mathrm{mts} \label{dist}\ee this is approximately the distance away from the decay region of the parent pion at which the final state leptons are produced, namely the ``decay vertex''. A detection of the charge lepton a distance $L$ away from the production region\emph{ may} mis-identify these charged leptons as being produced by neutrinos resulting from the decay of the parent meson or from oscillations between active neutrinos, the displaced decay vertex of the heavy sterile neutrino would then imply a \emph{shortening} the effective baseline by a large factor if $|U_{es}| \ll 1$.

Similar ``rare pion decay'' processes have been discussed within the context of lepton flavor violation in ref.\cite{cvetic} but without addressing the full time evolution.

\vspace{2mm}

\subsection{Entanglement and correlations.}

The asymptotic state after the parent particle and the intermediate resonant state have decayed is given by
\be |\Psi(\infty)\rangle =   \sum_{\vq,\vp}C_{\phi\chi\chi}(\vk,\vq,\vp;\infty)\,\big|\phi_{2,\vk-\vq}\chi_{1,\vp}\chi_{2,\vq-\vp} \rangle \,, \label{entan} \ee where
\be C_{\phi\chi\chi}(\vk,\vq,\vp;\infty) = \frac{\Mpi^*\, \Mfi^* \,e^{-i\epsfichi\,t}}{\Big[(\epsfichi-E^\pi)+i\frac{\Gamma_\pi}{2} \Big]\Big[(\epsfichi-\epsfi)+i\frac{\Gamma_{\phi_1}}{2} \Big]} \label{Cfichi2infi}\ee this is entangled state of three particles, which apart from $|\phi_{2,\vk-\vq}\rangle$ is very similar to the asymptotic entangled two-photon state from a   radiative cascade of two-level atomic systems in quantum optics\cite{qobooks1,mutu}.

Extrapolating the results to the case of   phenomenological relevance discussed above, the asymptotic state $|\Psi(\infty)\rangle$ is an entangled state of four particles. Quantum entanglement entails correlations, these are completely determined by the amplitudes $C(\infty)$, which depend on the mass and width of the resonant state and the matrix elements. These correlations will be manifest as intensity Hanbury-Brown-Twiss correlations among the charged lepton pairs produced by the decay of the resonant state that \emph{could} reveal important information on the properties of heavy sterile neutrinos in the intermediate state.

We postpone a deeper study of the time dependence of these   phenomenological consequences including mixing and oscillations to a future article.

\section{Conclusions and further questions:}
Cascade decays via resonant intermediate states are of interdsiciplinary interest, being ubiquitous in particle physics and quantum optics, and more recently are discussed within the context of inflationary cosmology. In this article we generalize methods of quantum optics to the realm of   quantum field theory to study the real time dynamics of cascade decay in a model quantum field theory of generic fields, however the conclusions are general.

The method is based on a hierarchical solution of the coupled equations for the amplitudes of the initial, intermediate and final multiparticle states. We show that a solution of the equations up to a given order in the interaction yields a non-perturbative resummation \emph{a la Dyson} in terms of the self-energies of the initial and intermediate resonant states.
We analyze the time evolution of the amplitudes and probabilities   and show that unitary time evolution is manifest  as a probability ``flow'' from the initial through the intermediate resonant and to the final state.
When the decay width of the initial parent particle $\Gamma_{\pi} $ is much larger than that of the intermediate resonant state $\Gamma_{\phi_1}$ there is a ``bottleneck'' in the evolution, the probability of the intermediate resonant state grows to a maximum nearly saturating unitarity on a time scale given
\be  t^* = \frac{\ln\Big[\frac{\Gamma_\pi}{\Gamma_{\phi_1}}\Big]}{\Gamma_\pi-\Gamma_{\phi_1}}  \ee and decays on a longer time scale $\simeq 1/\Gamma_{\phi_1}$, whereas in the opposite limit the population of the resonant state does not build up substantially and the probability flows almost directly from the initial to the final state on a time scale $1/\Gamma_{\pi}$.

We   provided an alternative formulation in terms of a quantum field theoretical generalization of the Wigner-Weisskopf method in quantum optics. This method provides a non-perturbative resummation of secular terms in time and reliably describes the time evolution of intermediate resonant states.  While both methods are equivalent, the quantum field theoretical generalization of  Wigner-Weisskopf   is directly applicable in the cosmological setting where the interaction Hamiltonian is explicitly dependent on time.

We \emph{conjecture} on potential phenomenological implications, in particular in the case of pseduscalar meson decay via  possible heavy sterile neutrinos as intermediate resonant states. Their production and decay into ``active'' neutrinos and charged leptons may have experimental relevance, we argue that cascade decay in this case may lead to ``displaced decay vertices'' which may result in  important corrections to the baseline dependence of the appearance (and dissapearance) probabilities. Asymptotically at long times after both the  parent and resonant intermediate particles decay, the final state  is a quantum entangled many particle state that features quantum correlations which are  completely determined by the asymptotic amplitudes. We also conjecture that quantum entanglement of the final state may translate in intensity Hanbury-Brown-Twiss correlations that may reveal information on the mass and width of the intermediate resonant state.

%%% new addition # 2 %%%%

The analysis and results presented above may be important in resonant leptogenesis\cite{lepto1,lepto2,lepto3,lepto4}, which is typically studied by implementing powerful non-equilibrium methods such as the Kadanoff-Baym and Keldysh formulations to obtain the kinetic description of the distribution functions that include resonant cross sections. However, the application  of the results obtained in this article to the important case of resonant leptogenesis is not direct: we obtained the evolution equations for the amplitudes of the many particle state resulting from the decay of an initial \emph{single particle state}. Instead   the Kadanoff-Baym and Keldysh approaches to resonant leptogenesis focus on the time evolution of \emph{distribution functions} which are ensemble averages in a non-equilibrium \emph{density matrix}. In order to apply the methods and results obtained above to this important case, first we must understand how to implement the non-perturbative methods described above to the case of a non-equilibrium density matrix rather than a single particle initial state.  The importance of resonant leptogenesis   motivates   further study to generalize the results obtained here to the case of non-equilibrium ensembles, correlation and distribution functions on which we expect to report in the future.

%%% end of new addition # 2 %%%%

\vspace{1mm}

\textbf{Further questions:}

In this article we focused on the description of the time evolution of cascade decay in a simple scenario with only one intermediate resonant state. An important case that remains to be studied is that of several intermediate resonant states that may result from mixing. In this case there will emerge interference phenomena manifest as oscillations, and if the intermediate states are nearly degenerate these oscillations and interference may lead to important dynamics at long times. This is the case for meson mixing relevant for CP and CPT violations. As discussed in the introduction cascade ``mixing'' is typically studied as   sequential events in (proper) time, however the analysis presented in the previous sections suggests that there may be   important corrections from the time dependence of the amplitudes that \emph{may} prove to be relevant to the experimental analysis. These questions along with an assessment of  potential impact on neutrino oscillation experiments from a resonant heavy  sterile neutrino, and the study of correlations in the final state as potential indicators of properties of the intermediate resonant state merit further study, which is postponed to a future article.

\vspace{1cm}

\textbf{Acknowledgements:}

 The authors acknowledge  partial support from NSF-PHY-1202227. D.B.  thanks P. Zoller and D. Jasnow for enlightening discussions.

\appendix

\section{Analysis of $C_{\phi\chi\chi}(\vk,\vq,\vp;t)$}\label{app:fintime}
To make the notation in this appendix more compact we introduce the following variables
\be \epsfichi-E^\pi_k = \eta ~~;~~
\epsfi-E^\pi_k = \sigma ~~;~~ \Delta = \Gamma_\pi-\Gamma_{\phi_1}~~;~~\Sigma = \Gamma_\pi+\Gamma_{\phi_1} \label{appdefs} \ee and suppressing the momenta labels, we write
\be C_{\phi\chi\chi}(t) \equiv D(t)+ C_{\phi\chi\chi}(\infty) \,.\label{ddef}\ee The contribution from the term $|C_{\phi\chi\chi}(\infty)|^2$ has been analyzed above, leading to eqn. (\ref{cinfi}), for the remaining terms we find
\be |D(t)|^2+ D(t) C^*_{\phi\chi\chi}(\infty)+D^*(t)C_{\phi\chi\chi}(\infty)= |\Mpi|^2~\, |\Mfi|^2 \Big[ (a)+(b)+(c)+(d)+(e) \Big]\, \label{dcs}\ee

 \be (a) = \frac{1}{\sigma^2+\Big(\frac{\Delta}{2}\Big)^2}\Bigg[\frac{ e^{-\Gamma_\pi\,t}- e^{-\frac{\Gamma_\pi}{2}\,t}(e^{i\eta\,t}+e^{-i\eta\,t}) }{\eta^2+\Big(\frac{\Gamma_\pi}{2}\Big)^2} \Bigg] \label{a}\ee

 \vspace{2mm}

 \be (b) = \frac{1}{\sigma^2+\Big(\frac{\Delta}{2}\Big)^2}\Bigg[\frac{ e^{-\Gamma_{\phi_1}\,t}- e^{-\frac{\Gamma_{\phi_1}}{2}\,t}(e^{i(\eta-\sigma)\,t}+e^{-i(\eta-\sigma)\,t}) }{(\eta-\sigma)^2+\Big(\frac{\Gamma_{\phi_1}}{2}\Big)^2} \Bigg] \label{b} \ee

\vspace{2mm}

\be (c) = \frac{1}{\sigma^2+\Big(\frac{\Delta}{2}\Big)^2}\Bigg[ \frac{ e^{i\eta\,t}\, e^{-\frac{\Gamma_{\pi}}{2}\,t}- e^{i\sigma\,t}\, e^{-\frac{\Sigma}{2}\,t} }{\Big(\eta +i\frac{\Gamma_{\pi}}{2}\Big)\,\Big(\eta-\sigma-i\frac{\Gamma_{\phi_1}}{2}\Big)}   \Bigg] \label{c} \ee

 \vspace{2mm}

\be (d) =  \frac{1}{\sigma^2+\Big(\frac{\Delta}{2}\Big)^2}\Bigg[ \frac{ e^{i(\eta-\sigma)\,t}\, e^{-\frac{\Gamma_{\phi_1}}{2}\,t}- e^{-i\sigma\,t}\, e^{-\frac{\Sigma}{2}\,t} }{\Big(\eta -i\frac{\Gamma_{\pi}}{2}\Big)\,\Big(\eta-\sigma+i\frac{\Gamma_{\phi_1}}{2}\Big)}   \Bigg] \label{d} \ee

 \vspace{2mm}

\be (e) =  \frac{1}{\sigma^2+\Big(\frac{\Delta}{2}\Big)^2}\Bigg[\frac{ e^{-i\eta\,t}\, e^{-\frac{\Gamma_{\pi}}{2}\,t}}{\Big(\eta -i\frac{\Gamma_{\pi}}{2}\Big)\,\Big(\eta-\sigma+i\frac{\Gamma_{\phi_1}}{2}\Big)}
+\frac{ e^{-i(\eta-\sigma)\,t}\, e^{-\frac{\Gamma_{\phi_1}}{2}\,t}}{\Big(\eta +i\frac{\Gamma_{\pi}}{2}\Big)\,\Big(\eta-\sigma-i\frac{\Gamma_{\phi_1}}{2}\Big)} \Bigg] \,.\label{e} \ee

The resonant denominators result in that the dominant contribution in the narrow widths limit are proportional to $\delta(\sigma)\delta(\eta)$, in order to extract the proportionality factors we integrate the above expressions in the complex $\sigma,\eta$ planes where the resonant denominators yield complex poles. We find
\be (a)+ (b) = -\frac{(2\pi)^2}{|\Delta|}~\Big\{ \frac{e^{- {\Gamma_{\pi}} \,t}}{\Gamma_{\pi}}~\delta(\eta) + \frac{e^{- {\Gamma_{\phi_1}} \,t}}{\Gamma_{\phi_1}}~\delta(\eta-\sigma)\Big\}~\delta(\sigma)\,. \label{apb}\ee
The integrals in the complex planes of $(c),(d)$ feature vanishing residues at the complex poles in $\eta,\eta-\sigma$, therefore these integrals yield subleading contributions in the narrow width limit.
Finally by the same procedure we find
\be (e) = 2 \frac{(2\pi)^2}{|\Delta|} ~~ \frac{e^{-\frac{|\Delta|}{2}\,t}~e^{-\frac{\Sigma}{2}\,t}}{\frac{1}{2}\,(|\Delta|+\Sigma)}~\delta(\eta)~\delta(\sigma)\,. \label{efin}\ee The final result is given by
\be (a)+(b)+(c)+(d)+(e) = \frac{(2\pi)^2}{\Gamma_\pi-\Gamma_{\phi_1}}~\Bigg\{
\frac{e^{- {\Gamma_{\pi}} \,t}}{\Gamma_{\pi}}- \frac{e^{- {\Gamma_{\phi_1}} \,t}}{\Gamma_{\phi_1}}\Bigg\}\delta(E^\pi_k-\epsfichi)~\delta(E^\pi_k-\epsfi) \label{totalsum}\ee

\vspace{3mm}

\section{The long time limit of eqn. (\ref{homocoeff})}\label{appB}

Introducing
\be \rho(\omega) = \sum_{\vq}  |\Mfi|^2 \delta(\omega-\epsfichi)\,,\label{rhoapB} \ee
we can write
\bea \int^t_0 dt' W^{\phi}_0(t',t')  & = &
 i\,t\,\int_{-\infty}^{\infty} d\omega' \, \frac{\rho(\omega')}{( \epsfi-\omega')}\,\Bigg[ 1-\frac{\sin(\omega'-\epsfi)t}{(\omega'-\epsfi)t} \Bigg] \nonumber \\ &  + &    \int_{-\infty}^{\infty} d\omega' \, \frac{\rho(\omega')}{( \epsfi-\omega')^2}\,\Bigg[ 1-\cos\big[(\omega'-\epsfi)t\big] \Bigg] \,. \label{energyminko}\eea

Asymptotically as $t\rightarrow \infty$, these integrals approach:
 \be \int_{-\infty}^{\infty} d\omega' \, \frac{\rho(\omega')}{( \epsfi-\omega')}\,\Bigg[ 1-\frac{\sin(\omega'-\epsfi)\,t}{(\omega'-\epsfi)\,t} \Bigg] ~~\overrightarrow{t\rightarrow \infty} ~~ \mathcal{P}\int_{-\infty}^{\infty} d\omega' \, \frac{\rho(\omega')}{( \epsfi-\omega')} \label{realpartofE}\ee
 \be \int_{-\infty}^{\infty} d\omega' \, \frac{\rho(\omega')}{( \epsfi-\omega')^2}\,\Bigg[ 1-\cos\big[(\omega'-\epsfi)t\big] \Bigg]~~ \overrightarrow{t\rightarrow \infty} ~~\pi\,t\, \rho(\epsfi) \,. \label{imagE}\ee

 The second integral above can be easily recognized as the  usual Fermi's Golden rule by taking the time derivative  .

\end{document}